\documentclass[aps,prd,showpacs,twocolumn,superscriptaddress,nofootinbib,preprintnumbers]{revtex4-1} 

\makeatletter
\def\p@subsection{}
\makeatother
\usepackage{color,graphicx,amsmath,amssymb,lipsum,bm}
\usepackage[colorlinks=true,citecolor=blue,linkcolor=blue,urlcolor=blue, backref=false,pdfborder={0 0 0}]{hyperref}
\usepackage[normalem]{ulem}
\usepackage[utf8]{inputenc} 
\usepackage{mathtools}
\usepackage{float}
\usepackage{ulem}
\usepackage{diagbox}
\usepackage[dvipsnames]{xcolor}
\usepackage{mathrsfs}
\usepackage{multirow}
\usepackage{amsmath}

\newcommand{\be}{\begin{equation}}
\newcommand{\ee}{\end{equation}}
\newcommand{\beqa}{\begin{eqnarray}}
\newcommand{\eeqa}{\end{eqnarray}}

\newcommand{\bseq}{\begin{subequations}}
\newcommand{\eseq}{\end{subequations}}

\def\gsim{\raise0.3ex\hbox{$\;>$\kern-0.75em\raise-1.1ex\hbox{$\sim\;$}}}
\def\lsim{\raise0.3ex\hbox{$\;<$\kern-0.75em\raise-1.1ex\hbox{$\sim\;$}}}
\def\beqn#1{\begin{equation}\label{#1}}
\def\eeqn{\end{equation}}
\def\beqa#1{\begin{eqnarray}\label{#1}}
\def\eeqa{\end{eqnarray}}

\def\Z2{$\mathcal{Z_2}$}

\newcommand {\ignore}[1]{}

\begin{document}

\title{DDELenS: Tomographic Weak Lensing Simulations with Dynamical Dark Energy}

\author{Shu-Fan Chen} 
\email{sc5848@columbia.edu
 } 
\affiliation{Astrophysics Laboratory, Columbia University, New York, NY 10027, USA}
\affiliation{Department of Physics, Columbia University, New York, NY 10027, USA}
\affiliation{Department of Astronomy, Columbia University New York, NY 10027, USA}

\author{J.~Colin Hill}
\affiliation{Astrophysics Laboratory, Columbia University, New York, NY 10027, USA}
\affiliation{Department of Physics, Columbia University, New York, NY 10027, USA}

\author{Zolt\'an Haiman}
\affiliation{Institute of Science and Technology Austria (ISTA), Am Campus 1, Klosterneuburg 3400, Austria}
\affiliation{Department of Astronomy \& Astrophysics, Columbia University New York, NY 10027, USA}
\affiliation{Department of Physics, Columbia University, New York, NY 10027, USA}

\begin{abstract}
We present a suite of 128 cosmological simulations in $w_0w_a$CDM cosmologies with massive neutrinos, designed to enable simulation-based inference of dynamical dark energy parameters from weak gravitational lensing and other cosmological observables. Each simulation follows $1024^3$ dark matter particles in a $512\,h^{-1}$Mpc periodic box, with cosmological parameters sampled using a Sobol sequence over the ranges $\Omega_m \in [0.28, 0.35]$, $\sigma_8 \in [0.74, 0.90]$, $w_0 \in [-1.0, -0.5]$, and $w_a \in [-1.5, 0.0]$, motivated by recent results from the Dark Energy Spectroscopic Instrument survey. The nonlinear evolution is computed with \texttt{PKDGRAV3} using the N-body gauge to consistently incorporate relativistic corrections from massive neutrinos, photons, and dark energy. For each cosmology, we generate 10,000 ray-traced convergence maps at six source redshifts $z_s \in \{0.5, 1.0, 1.5, 2.0, 2.5, 1100\}$ using \texttt{LensTools}, and additionally provide halo catalogues and projected halo overdensity fields for galaxy--lensing cross-correlation analyses. We generate 100,000 realizations at the fiducial cosmology for the construction of the covariance matrix. This simulation suite provides a foundation for field-level inference and non-Gaussian statistics with next-generation weak lensing surveys such as the Rubin Observatory LSST, Euclid, and the Nancy Grace Roman Space Telescope, as well as CMB lensing experiments including the Simons Observatory.
\end{abstract}

\maketitle

\section{Introduction}
The nature of dark energy remains one of the central open questions in cosmology. The first data release of the Dark Energy Spectroscopic Instrument (DESI DR1), combining baryon acoustic oscillation (BAO)~\cite{DESI:2024mwx} and full-shape power spectrum measurements~\cite{DESI:2024jxi,DESI:2024hhd} from galaxy samples spanning $z\in[0.1,2.1]$, found a mild but statistically inconclusive preference for dynamical dark energy --- specifically $w_0>-1$ and $w_a<0$ in the Chevallier--Polarski--Linder (CPL) parametrization --- when combined with cosmic microwave background (CMB) data. This preference strengthens upon inclusion of Type Ia supernova (SNe Ia) datasets~\cite{Brout:2022vxf,Rubin:2023jdq,DES:2024jxu,DES:2025sig}, some of which already show independent hints of departures from a cosmological constant. With two additional years of data, BAO measurements from DESI DR2~\cite{DESI:2025zpo,DESI:2025zgx} further consolidate this preference even without external datasets. More recently, a joint analysis of supernovae, BAO, weak lensing, and galaxy clustering from the full Dark Energy Survey (DES) dataset finds a $2.2\sigma$ preference for dynamical dark energy, increasing to $3.0\sigma$ when combined with DESI DR2 BAO and CMB data. A $2.6\sigma$ preference still persists when supernovae are excluded from the combined analysis, providing a cross-check independent of supernova photometric calibration~\cite{DES:2026jmi}.

Whether this signal reflects genuine new physics, however, remains an open question. On the supernova side, distance measurements are subject to ongoing systematic concerns, including redshift calibration uncertainties, foreground modeling, and host-galaxy environmental effects~\cite{Gialamas:2024lyw,Dhawan:2024gqy}. On the large-scale structure side, combining DESI with full-shape information from the low-redshift Baryon Oscillation Spectroscopic Survey (BOSS) weakens or eliminates the preference for dynamical dark energy even when SNe Ia data are included~\cite{Chen:2024vuf,Chen:2025jnr} (see however Ref.~\cite{Silva:2025twg}). Resolving this tension will require both more precise data and more accurate theoretical modeling across all scales.

Accurately predicting the clustering of matter in dynamical dark energy cosmologies is itself a significant theoretical challenge. While Einstein--Boltzmann solvers such as \texttt{CAMB}~\cite{Lewis:1999bs} and \texttt{CLASS}~\cite{Lesgourgues:2011re,Blas:2011rf} provide highly accurate large-scale predictions, they break down in the nonlinear regime. Fitting functions such as Halofit~\cite{Takahashi:2012em,Bird:2011rb} and HMCode~\cite{Mead:2020vgs} extend these to smaller scales, but they are insufficiently accurate for next-generation surveys. A complete description ultimately requires N-body simulations that evolve all species simultaneously\footnote{We do not include the nonlinear clustering of dark energy perturbations in our simulations as they are subdominant relative to matter fluctuations~\cite{Sefusatti:2011cm,Creminelli:2009mu}}.

Several weak lensing simulation suites have been developed to study non-Gaussian statistics and beyond-power-spectrum constraints in non-standard cosmologies. For example, early work~\cite{Dietrich:2009jq,Gupta:2018eev,ZorrillaMatilla:2016vae,Kratochvil:2009wh} explored different non-Gaussian statistics such as peak statistics to study $\Omega_m$ and $\sigma_8$; MassiveNuS~\cite{Liu:2017now} extended this to massive neutrinos; CosmoGridV1~\cite{Kacprzak:2022pww,Fluri:2022rvb} sampled 2500 points across a six-parameter $w$CDM space; Ulagam~\cite{Anbajagane:2023wif} probes primordial non-Gaussianity; HalfDome~\cite{Bayer:2024egi} and Takahashi+~\cite{Takahashi:2017hjr} include correlated CMB and large-scale structure (LSS) mocks.

On the emulator side, a number of efforts --- including COYOTE~\cite{Heitmann:2008eq,Heitmann:2009cu,Lawrence:2009uk}, EuclidEmulator2~\cite{Euclid:2020rfv}, BACCO~\cite{Angulo:2020vky}, CAMELUS~\cite{Lin:2014dua}, CosmicEmu~\cite{Lawrence:2017ost}, CSST~\cite{Chen:2025ugn,Chen:2025opy,Zhou:2025iiu}, Aemulus~\cite{DeRose:2023dmk,DeRose:2018xdj,McClintock:2018uyf,Zhai:2018plk}, and Goku~\cite{Yang:2025dtc} --- have provided accurate nonlinear predictions of matter power spectra for beyond-$\Lambda$CDM cosmologies, though these have largely focused on the matter power spectrum or halo mass function rather than weak lensing maps. Approximate methods such as COmoving Lagrangian Acceleration (COLA) have also been employed for emulator construction~\cite{Reboucas:2025hue}. On the simulation methodology side, Nefertiti represents the first hydrodynamical simulation to self-consistently incorporate nonlinear dark energy clustering~\cite{Blot:2026var,Blot:2022mnt}. In this work, we employ a pipeline combining \texttt{PKDGRAV3}~\cite{Potter:2016ttn} and \texttt{CONCEPT}~\cite{Dakin:2019dxu, Dakin:2019vnj}, in which dark energy perturbations are evolved linearly while gravity is solved at full N-body precision, offering an efficient and self-consistent framework for dynamical dark energy simulations.

Despite this progress, no existing weak lensing simulation suite covers dynamical dark energy in the $w_0w_a$CDM framework, motivated by recent DESI results, with the full machinery needed for field-level and non-Gaussian inference. Current surveys such as DES~\cite{DES:2026mkc,DES:2026fyc}, KiDS~\cite{Wright:2025xka}, and HSC~\cite{Miyatake:2023njf} have already demonstrated the constraining power of galaxy weak lensing beyond the power spectrum and constraints on $w_0$ from some of them. CMB lensing measurements, for example from ACT~\cite{ACT:2023dou,ACT:2023kun}, SPT~\cite{Wu:2019hek}, SPT-3G~\cite{SPT:2023jql}, Planck~\cite{Planck:2018lbu,Carron:2022eyg}, and their combination~\cite{ACT:2025qjh}, have also provided cosmological constraints through the lensing convergence power spectrum (see Refs.~\cite{Liu:2016nfs,Chen:2026jlr} for forecasts beyond the power spectrum, for example). Next-generation experiments --- including the Rubin Observatory LSST~\cite{LSSTDarkEnergyScience:2018yem}, the Nancy Grace Roman Space Telescope~\cite{Akeson:2019biv,Spergel:2015sza}, Euclid~\cite{Euclid:2020rfv}, and CMB lensing from the Simons Observatory~\cite{SimonsObservatory:2018koc,SimonsObservatory:2025wwn} --- will deliver datasets of sufficient statistical precision to demand accurate, field-level forward models. In this work, we present such a suite. We combine \texttt{PKDGRAV3}~\cite{Potter:2016ttn} and \texttt{CONCEPT}~\cite{Dakin:2019dxu, Dakin:2019vnj} for nonlinear N-body evolution --- incorporating relativistic corrections from all linear species via the N-body gauge --- with \texttt{LensTools}~\cite{Petri:2016zrq} for fully ray-traced convergence maps and \texttt{ROCKSTAR}~\cite{Behroozi:2011ju} for halo finding. We generate 10,000 ray-traced convergence maps at six source redshifts for each of 128 cosmologies sampled with a Sobol sequence over $(w_0, w_a, \Omega_m, \sigma_8)$, along with halo catalogues and projected halo overdensity fields suitable for galaxy--lensing cross-correlation analyses.

This paper is organized as follows. Section~\ref{sec:theoreticalBackground} reviews the theoretical background, covering the dark energy parametrization, N-body gauge, and weak gravitational lensing. Section~\ref{sec:cosmologicalSimulation} describes the simulation pipeline. We present validation of the matter power spectrum in Section~\ref{sec:matterPowerSpectrum}, the halo catalogue and mass function in Section~\ref{sec:haloCatalogue}, and the lensing convergence maps and power spectra in Section~\ref{sec:lensingConvergenceMaps}. We conclude in Section~\ref{sec:conclusion}.

\section{Theoretical Background}\label{sec:theoreticalBackground}
Our simulations are run in a $w_0 w_a$CDM cosmology with massive neutrinos. We briefly summarize below the two theoretical ingredients most relevant to our simulation setup.

\subsection{Dynamical Dark Energy}
We model the dark energy equation of state using the Chevallier--Polarski--Linder (CPL) parameterization~\cite{Chevallier:2000qy, Linder:2002et},
\begin{align}
    w(a) = w_0 + w_a(1-a),
\end{align}
where $a$ is the cosmic scale factor. When $w$ crosses $-1$, the standard fluid description breaks down due to a divergence in the $(1+w)^{-1}$ factor in the dark energy Euler equation. To handle such phantom-crossing cosmologies, we adopt the parametrized post-Friedmann (PPF) formalism~\cite{Baker:2012zs}, which closes the perturbation equations by relating the dark energy momentum density to that of dark matter on large scales, thereby preserving energy-momentum conservation across the phantom divide.

\subsection{N-body Gauge}
Newtonian N-body simulations must be treated with care when multiple linear species are present, as the standard Newtonian framework is not straightforwardly consistent with general relativity. In our simulations, we therefore adopt the N-body gauge (Nb), in which the equations of motion are augmented by a linear relativistic correction potential $\gamma^{\rm Nb}$~\cite{Fidler:2016tir,Fidler:2017ebh} implemented by the combination of \texttt{PKDGRAV3} and \texttt{CONCEPT} (see Section~\ref{sec:cosmologicalSimulation} for more details):
\begin{align}
    \left(\partial_\tau + \mathcal{H}\right)
    \mathbf{v}^{\rm Nb}_{{\rm CDM}+b} 
    &= -\nabla\phi + \nabla\gamma^{\rm Nb}, \\
    \nabla^2 \phi 
    &= 4\pi G a^2\, \delta\rho^{\rm Nb}_{\rm tot},
\end{align}
where $\tau$ is conformal time, $\mathcal{H} = aH$ is the conformal Hubble parameter, $\phi$ is the gravitational potential, and
\begin{align}
    \delta\rho^{\rm Nb}_{\rm tot} = \delta\rho_{\rm CDM} + \delta\rho_{b} 
    + \delta\rho_{\gamma} + \delta\rho_{\nu} + \delta\rho_{\rm DE}
\end{align}
is the total density perturbation summed over all species. The correction potential $\gamma^{\rm Nb}$ encodes the momentum density, pressure, and anisotropic stress of all relativistic species not directly tracked as particles in the simulation, and satisfies
\begin{align}
    \nabla^2 \gamma^{\rm Nb} = -4\pi G a^2\, \delta\rho_{\rm metric},
\end{align}
where $\delta\rho_{\rm metric}$ is a fictitious source whose Newtonian gravity captures all relevant general relativistic effects. As a result, the simulation output can be gauge-transformed to any observationally relevant gauge in post-processing at linear order.

\section{Cosmological Simulations}\label{sec:cosmologicalSimulation}

We now describe the simulation pipeline used to generate our suite of $w_0w_a$CDM cosmologies.

\subsection{Initial Conditions}

Initial conditions are generated using second-order Lagrangian perturbation theory (2LPT) as implemented in \texttt{NGENIC}~\cite{Springel:2005mi}, with the linear matter power spectrum computed by \texttt{CLASS}~\cite{Blas:2011rf}\footnote{Code and documentation can be found in \url{http://class-code.net/}} at $z = 49$. All cosmologies share the same initial random seed to ensure that differences between simulations reflect cosmological parameter variations rather than sample variance. For the fiducial cosmology, we additionally generate three independent realizations to better estimate the covariance matrix.

\subsection{Pre-computation of Transfer Functions}

Prior to running the N-body simulations, all cosmological background and linear perturbation quantities are precomputed using \texttt{CONCEPT}~\cite{Dakin:2019dxu, Dakin:2019vnj}\footnote{Code and documentation can be found in \url{https://github.com/jmd-dk/concept/} and \url{https://jmd-dk.github.io/concept/index.html}, respectively.}
interfaced with \texttt{CLASS}. \texttt{CONCEPT} takes the synchronous-gauge output from \texttt{CLASS} and converts it to the N-body gauge, yielding transfer functions for all relevant species — CDM$+$baryons, photons, massive neutrinos, dark energy, and the metric perturbations — via the gauge transformation
\begin{align}
    \delta\rho^{\rm Nb}_\alpha &= \delta\rho^{\rm s}_\alpha + 3\mathcal{H}(1 + w_\alpha)\bar{\rho}_\alpha \frac{\theta^{\rm s}_{\rm tot}}{k^2}, \label{eq:gauge_rho} \\
    \theta^{\rm Nb}_\alpha &= \theta^{\rm s}_\alpha + \partial_\tau\!\left(\frac{h_{\rm mp}}{2} - 3\mathcal{H}\frac{\theta^{\rm s}_{\rm tot}}{k^2}\right), \label{eq:gauge_theta}
\end{align}
where superscripts ${\rm Nb}$ and ${\rm s}$ denote the N-body and synchronous gauges respectively, $\alpha$ labels the species, $h_{\rm mp}$ is the trace of the metric perturbation in synchronous gauge, and $\theta_{\rm tot}$ is the total velocity divergence of all species. The resulting transfer functions, together with background quantities such as $H(a)$ that are non-trivial to compute in the presence of massive neutrinos, are stored on a global $(a, k)$ grid in an HDF5 file and read in by \texttt{PKDGRAV3} at runtime. For the resolution of the tabulated data, we have 256 modes within $k\in[10^{-3}, 15]$ Mpc$^{-1}$ and 512 time stamps in the range of $z\in[0,50]$\footnote{An example script that includes massive neutrinos can be found in \url{https://jmd-dk.github.io/concept/utilities/class.html\#complete-example-massive-neutrinos}. To include the dynamical dark energy, we can simply update the \texttt{CLASS} parameters by adding \texttt{w0\_fld} and \texttt{wa\_fld}.}.

\subsection{Nonlinear Evolution}\label{sec:nonlinearEvolution}

The nonlinear evolution of the dark matter particles is computed with \texttt{PKDGRAV3}~\cite{Potter:2016ttn}\footnote{Code and documentation can be found in \url{https://bitbucket.org/dpotter/pkdgrav3/src/master/} and \url{https://pkdgrav3.readthedocs.io/en/latest/}, respectively.}
using a fast multipole method (FMM)~\cite{1987JCoPh..73..325G, 2002JCoPh.179...27D} combined with adaptive multi-timestepping. At each base time step, \texttt{PKDGRAV3} reads the precomputed transfer functions from the HDF5 file, converts the linear species — massive neutrinos, photons, dark energy, and the metric field — into density fields on a grid, and applies them as a weak corrective mesh force to the dark matter particles. This ensures that the nonlinear evolution of the dark matter field self-consistently accounts for all linear species.

We use $N_{\rm part} = 1024^3$ dark matter particles in a periodic box of side length $512\,h^{-1}$Mpc, giving a particle mass of $M_{\rm part} = \rho_{\rm c} V \Omega_m / N_{\rm part} = \Omega_m \times 3.48 \times 10^{10}\,h^{-1}M_\odot$, depending on the cosmology. Linear species are evolved on a $512^3$ grid using the pre-computed tabular data. Snapshots are saved at comoving intervals of $180\,h^{-1}$Mpc from $z = 0$ to $z = 49$, consistent with the spacing adopted by MassiveNuS~\cite{Liu:2017now}, and sufficient for ray-tracing to the CMB last-scattering surface. Snapshots are saved at the comoving intervals of $180\,h^{-1}$Mpc by precomputing, for each cosmology, the corresponding output redshifts using \texttt{CLASS} and passing them to \texttt{PKDGRAV3} via the \texttt{dRedTo} parameter, with \texttt{nSteps}\,$=$\,\texttt{iOutInterval}.

\subsection{Parameter Sampling}

\begin{figure*}[t]
    \centering
    \includegraphics[width=0.7\textwidth]{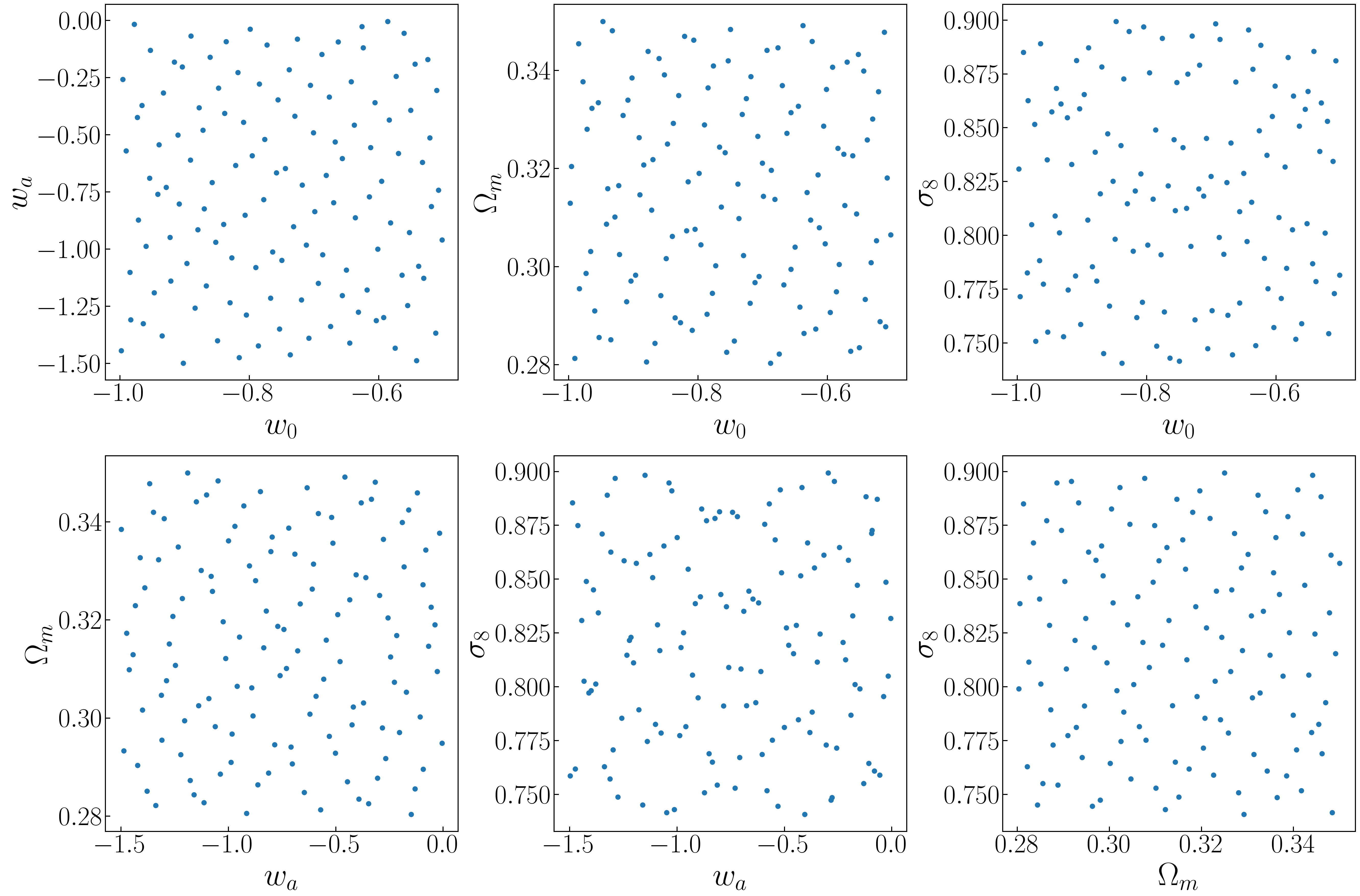}
    \caption{Pairwise distribution of the 128 dark energy cosmologies sampled using a Sobol sequence in the four-dimensional parameter space $(w_0, w_a, \Omega_m, \sigma_8)$, with ranges $w_0 \in [-1.0, -0.5]$, $w_a \in [-1.5, 0.0]$, $\Omega_m \in [0.28, 0.35]$, and $\sigma_8 \in [0.74, 0.90]$.}
    \label{fig:parameter}
\end{figure*}

We sample four cosmological parameters — $\Omega_m$, $\sigma_8$, $w_0$, and $w_a$ — across 128 cosmologies using a Sobol sequence~\cite{Sobol1967OnTD}, which provides quasi-random coverage of the parameter space with better uniformity than a Latin hypercube. We assume a flat universe ($\Omega_K = 0$) and fix the remaining parameters to $\Omega_b = 0.04851$, $h = 0.684$, $n_s = 0.9751$, and $\sum_i m_{\nu_i} = 0.06\,\mathrm{eV}$, motivated by the ACT DR6 cosmological constraints~\cite{AtacamaCosmologyTelescope:2025blo,AtacamaCosmologyTelescope:2025nti}. The three neutrino mass eigenstates are assumed degenerate and evolved linearly. Our prior ranges,
\begin{alignat}{2}
    &\Omega_m \in [0.28,\,0.35]\,,\quad &\sigma_8 \in [0.74,\,0.90]\,, \nonumber\\
    &w_0 \in [-1.0,\,-0.5]\,,\quad &w_a \in [-1.5,\,0.0]\,,
\end{alignat}
are motivated by recent DESI DR1 and DR2 results~\cite{DESI:2024hhd,DESI:2025zgx}. The fiducial cosmology, used to estimate the covariance matrix, is set to $\Omega_m = 0.300$, $\sigma_8 = 0.8117$, $w_0 = -1$, and $w_a = 0$. The pairwise parameter distribution is shown in Figure~\ref{fig:parameter}. We also add an additional parameter set at the center of this prior range.

\section{Matter Power Spectrum}\label{sec:matterPowerSpectrum}
In this section, we show the matter power spectrum for each cosmology at $z=0$. We use \texttt{Pylians3}~\cite{Pylians}\footnote{Code is publicly available in \url{https://pylians3.readthedocs.io/en/master/}} to first create a mesh with $1024^3$ voxels using the triangular-shaped-cloud (TSC) scheme, then we compute the matter power spectrum from it. 

\begin{figure}[t]
    \centering
    \includegraphics[width=0.85\linewidth]{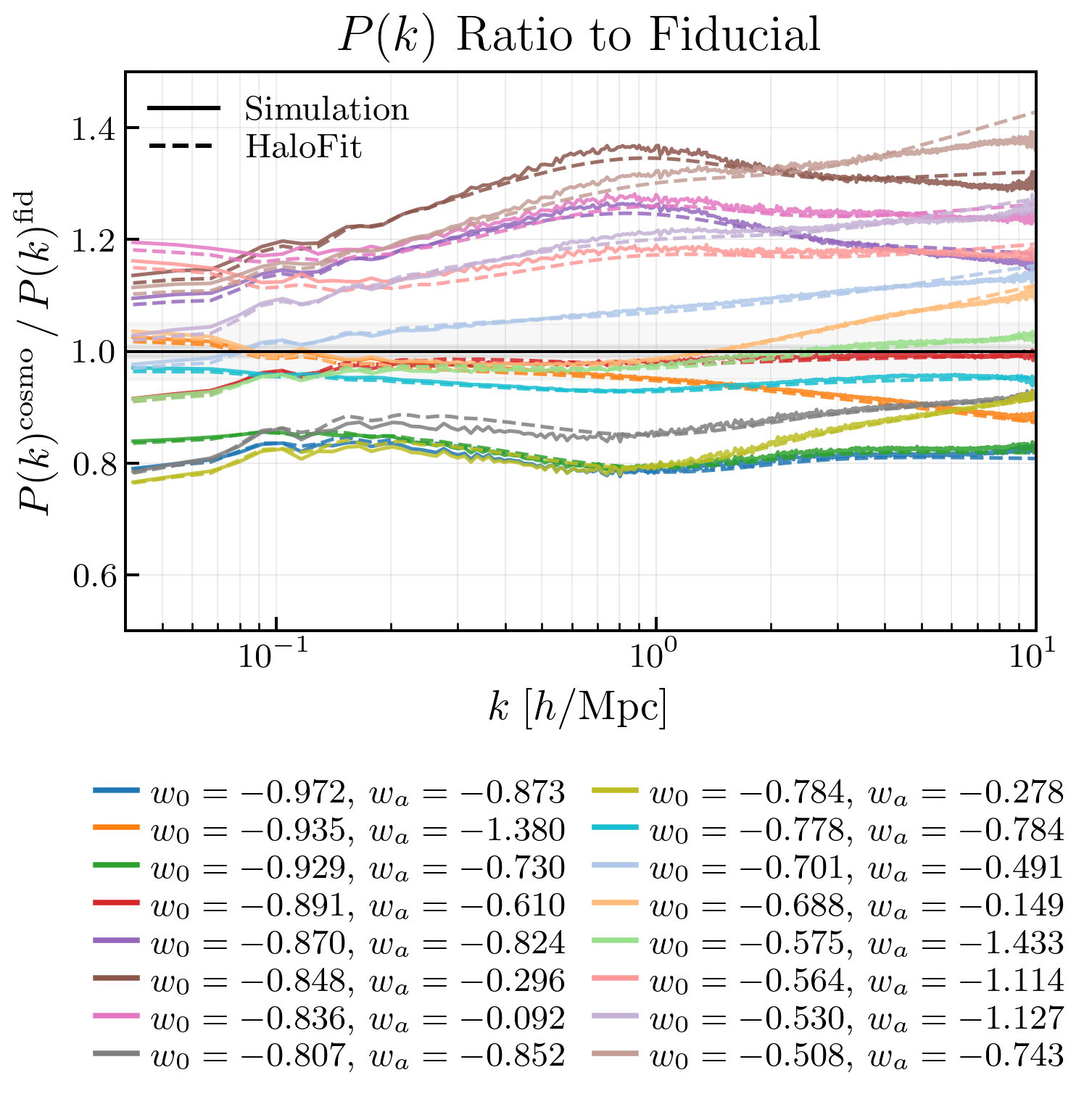}
    \caption{Ratio of the matter power spectrum $P(k)$ to the fiducial $\Lambda$CDM cosmology at $z=0$, for 16 randomly selected $w_0w_a$CDM dark energy cosmologies. Solid lines show simulation results and dashed lines show the corresponding HMcode-2020 predictions.}
    \label{fig:pk_ratio}
\end{figure}

As a validation of our N-body simulations, we compare the measured matter power spectrum $P(k)$ against the HMcode-2020~\cite{Mead:2020vgs} predictions across our suite of dark energy cosmologies at $z=0$. Figure~\ref{fig:pk_ratio} shows the ratio of $P(k)$ to the fiducial $\Lambda$CDM cosmology at $z=0$ for 16 randomly selected $w_0w_a$CDM cosmologies. Overall, we find good agreement between the simulation measurements and the HMcode-2020 predictions, with residual deviations generally within $\sim 5\%$ on quasi-linear scales. Larger deviations at small scales ($k \gtrsim 1\,h\,\mathrm{Mpc}^{-1}$) are expected given the known limitations of fitting functions in the highly nonlinear regime, as well as the finite mass resolution of our simulations. 

\section{Halo Catalogue}\label{sec:haloCatalogue}

We use \texttt{ROCKSTAR}~\cite{Behroozi:2011ju}\footnote{Code is publicly available in \url{https://bitbucket.org/gfcstanford/rockstar/src/main/}} as our halo finder. \texttt{ROCKSTAR} is a phase-space friends-of-friends halo finder that identifies halos and subhalos by grouping particles in six-dimensional phase space, making it particularly robust against spurious halo detections in dense environments. For each snapshot in our simulation suite, we run \texttt{ROCKSTAR} on the particle output of \texttt{PKDGRAV3} to identify all gravitationally bound structures and save them into .ascii files. We set the linking length to 0.28 as default and retain all halos with a minimum of 10 particles, which ensures that the internal structure of retained halos is reasonably resolved except for the smallest halos. Halo masses are computed using three definitions: $M_{\rm vir}$, $M_{\rm 200b}$, and $M_{\rm 200c}$\footnote{$M_{\rm vir}$ is the mass enclosed within the virial radius $r_{\rm vir}$, defined using the spherical overdensity criterion of Ref.~\cite{1998ApJ...495...80B}, in which the mean enclosed density equals $\Delta_{\rm vir}(z)\,\rho_{\rm crit}(z)$, with $\Delta_{\rm vir}(z)$ derived from the spherical collapse model and depending on cosmology and redshift. $M_{\rm 200b}$ is instead the mass enclosed within a radius $r_{\rm 200b}$ inside which the mean density is $200$ times the mean matter density of the universe, $\rho_m(z)$, while $M_{\rm 200c}$ is defined analogously but with respect to the critical density $\rho_{\rm crit}(z)$, such that the mean enclosed density within $r_{\rm 200c}$ equals $200\,\rho_{\rm crit}(z)$.}, to facilitate comparison with theoretical predictions and other simulation suites. Host and subhalo relationship is then determined in post-processing using a Python reimplementation of the \texttt{ROCKSTAR} parent-finding algorithm\footnote{The parent-finding algorithm is implemented in \texttt{parent.c} within the \texttt{ROCKSTAR} source code.}, which assigns each halo a parent based on a radial overlap criterion.

As a validation of our halo finding pipeline, we compare the measured halo mass function (HMF) against the theoretical prediction of Tinker et al.~(2008)~\cite{Tinker:2008ff}\footnote{We use \texttt{Colossus}~\cite{Diemer:2017bwl} to compute theoretical halo mass functions, which is publicly available in \url{https://bdiemer.bitbucket.io/colossus/index.html}. The dynamical dark energy only enters through the conversion of the overdensity threshold.} across a range of dynamical dark energy cosmologies, for all three mass definitions: $M_\mathrm{vir}$, $M_\mathrm{200b}$, and $M_\mathrm{200c}$.

\begin{figure*}[t]
    \centering
    \includegraphics[width=1.0\textwidth]{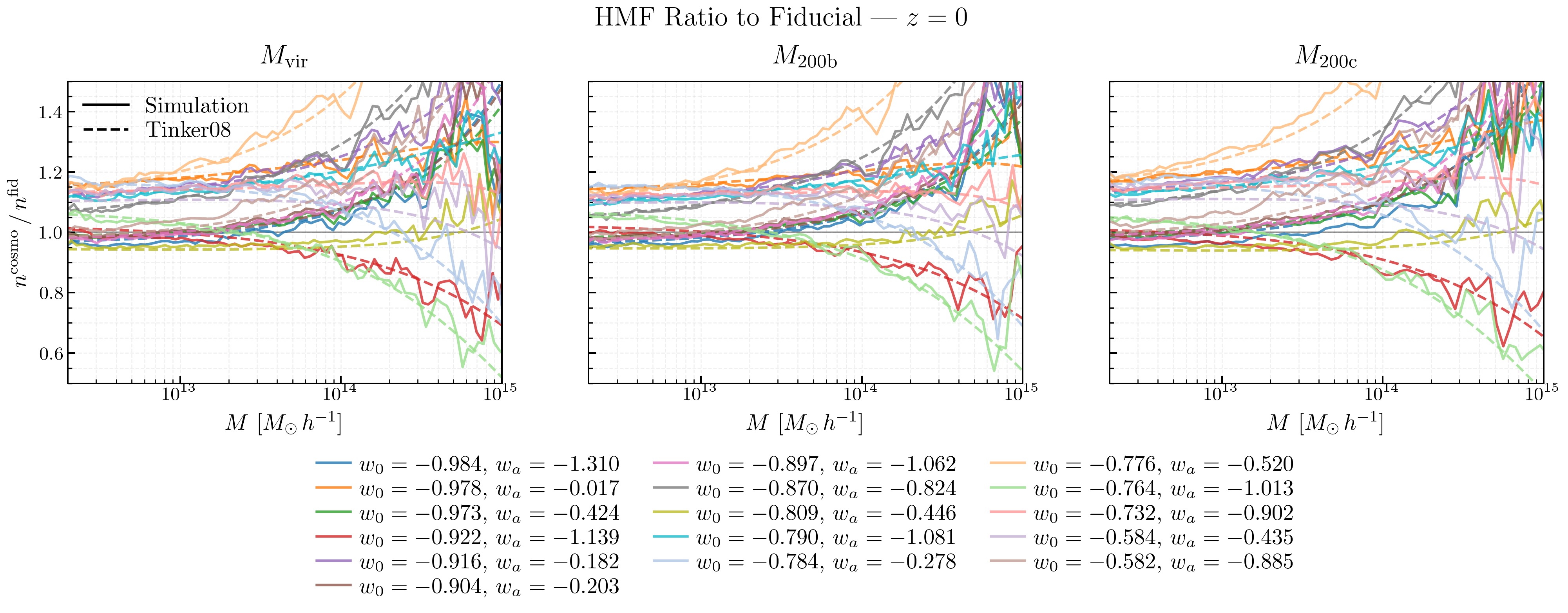}
    \caption{Ratio of the halo mass function (HMF) $n(M)$ to the fiducial $\Lambda$CDM cosmology at $z=0$, for 16 randomly selected $w_0$--$w_a$ dark energy cosmologies, shown for three mass definitions: $M_\mathrm{vir}$ (Left), $M_\mathrm{200b}$ (Center), and $M_\mathrm{200c}$ (Right). Solid lines show simulation results (smoothed) and dashed lines show the corresponding Tinker et al.~(2008) predictions.}
    \label{fig:hmf_ratio}
\end{figure*}

Figure~\ref{fig:hmf_ratio} shows the ratio of the HMF to the fiducial $\Lambda$CDM cosmology at $z=0$ for 16 randomly selected $w_0$-$w_a$ cosmologies. The Tinker et al.~(2008) fitting function is known to reproduce simulation results to within a few percent accuracy across a wide range of cosmologies and redshifts, making it a robust benchmark for validation. Overall, we find good agreement between our measured HMFs and the theoretical predictions across all three mass definitions, with deviations generally within the $\pm 10\%$ band. The level of agreement is comparable across $M_\mathrm{vir}$, $M_\mathrm{200b}$, and $M_\mathrm{200c}$, confirming that our halo finder and mass assignment are consistent regardless of the chosen mass definition. The residual scatter reflects both the theoretical error from the fitting function and the Poisson shot noise because of the finite sample size.

\begin{figure}[t]
    \centering
    \includegraphics[width=0.85\linewidth]{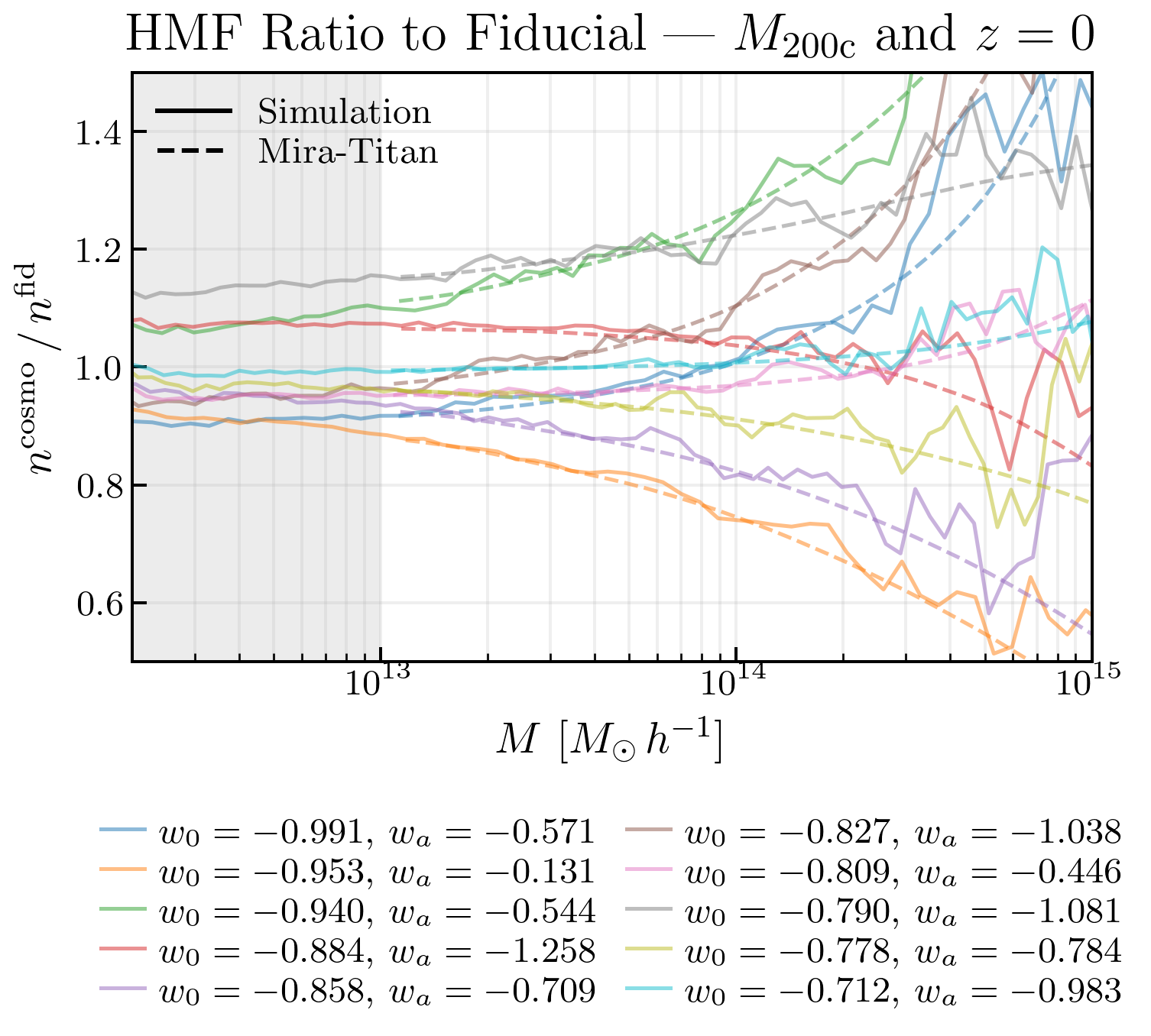}
    \caption{Ratio of the halo mass function (HMF) $n(M)$ to the fiducial $\Lambda$CDM cosmology at $z=0$, for 10 randomly selected $w_0$--$w_a$ dark energy cosmologies, over $M_\mathrm{200c}$. Solid lines show simulation results (smoothed) and dashed lines show the corresponding {\tt MiraTitanHMFemulator} predictions. The gray region indicates the mass range that is not available for the emulator.}
    \label{fig:hmf_ratio_miratitan}
\end{figure}

We also compare the ratio with the {\tt MiraTitanHMFemulator}~\cite{Bocquet:2020tes}\footnote{The emulator can be found in \url{https://miratitanhmfemulator.readthedocs.io/en/latest/index.html\#}} in Figure~\ref{fig:hmf_ratio_miratitan}. We show the comparison with 10 cosmologies that are within the training range of the emulator, and it shows similar agreement as the Tinker et al.~(2008) halo mass function. 


\section{Lensing Convergence Maps}\label{sec:lensingConvergenceMaps}
We use \texttt{LensTools}~\cite{Petri:2016zrq}\footnote{Code and documentation can be found in \url{https://github.com/apetri/LensTools} and \url{https://lenstools.readthedocs.io/en/latest/}, respectively. It is important to follow the steps to construct the directory tree and have all relevant parameters in the name of the directory, including $\Omega_m$, $\sigma_8$, $w_0$, and $w_a$.} to run full ray-tracing through the simulations\footnote{Using only Born approximation can lead to biased estimation for higher-order statistics for both low- and high-redshift lensing~\cite{Petri:2016qya,Fabbian:2017wfp,Pratten:2016dsm}.}. As mentioned in Section~\ref{sec:nonlinearEvolution}, we have saved the snapshots every 180 Mpc from $z=0$ up until $z=49$ so that we can run ray-tracing to generate weak lensing convergence maps. To produce the lens planes, we cut each snapshot into four planes from each of the three directions, with a thickness of 180 Mpc per plane, yielding 12 planes with $4096^2$ pixels per snapshot\footnote{Before cutting the snapshots into multiple lens planes, we first need to transform the usual \texttt{Gadget}/HDF5 files to \texttt{LensTools}-style \texttt{Gadget} files using the class \texttt{GadgetSnapshotDE()}. The number of cores used to cut the plane should be at least the number of files of each snapshot.}. When we run ray-tracing, we randomly select one lens plane for each redshift and apply a random two-dimensional shift to improve the statistical independence across different realizations. This procedure has been shown to be valid to generate $\mathcal{O}(10^4)$ statistically independent weak lensing maps from a single N-body simulation~\cite{Petri:2016wlu}. We then shoot $2048^2$ light rays from the center of the map back to the source plane. After all of these steps, we downgrade the map to have $512^2$ pixels in total by averaging over the neighboring pixels. We have also modified and ensured that $\texttt{LensTools}$ uses the $w_0w_a$CDM cosmology together with the massive neutrino included as default to compute all the distances relevant to ray-tracing.

\begin{figure*}[t]
    \centering
    \includegraphics[width=0.75\textwidth]{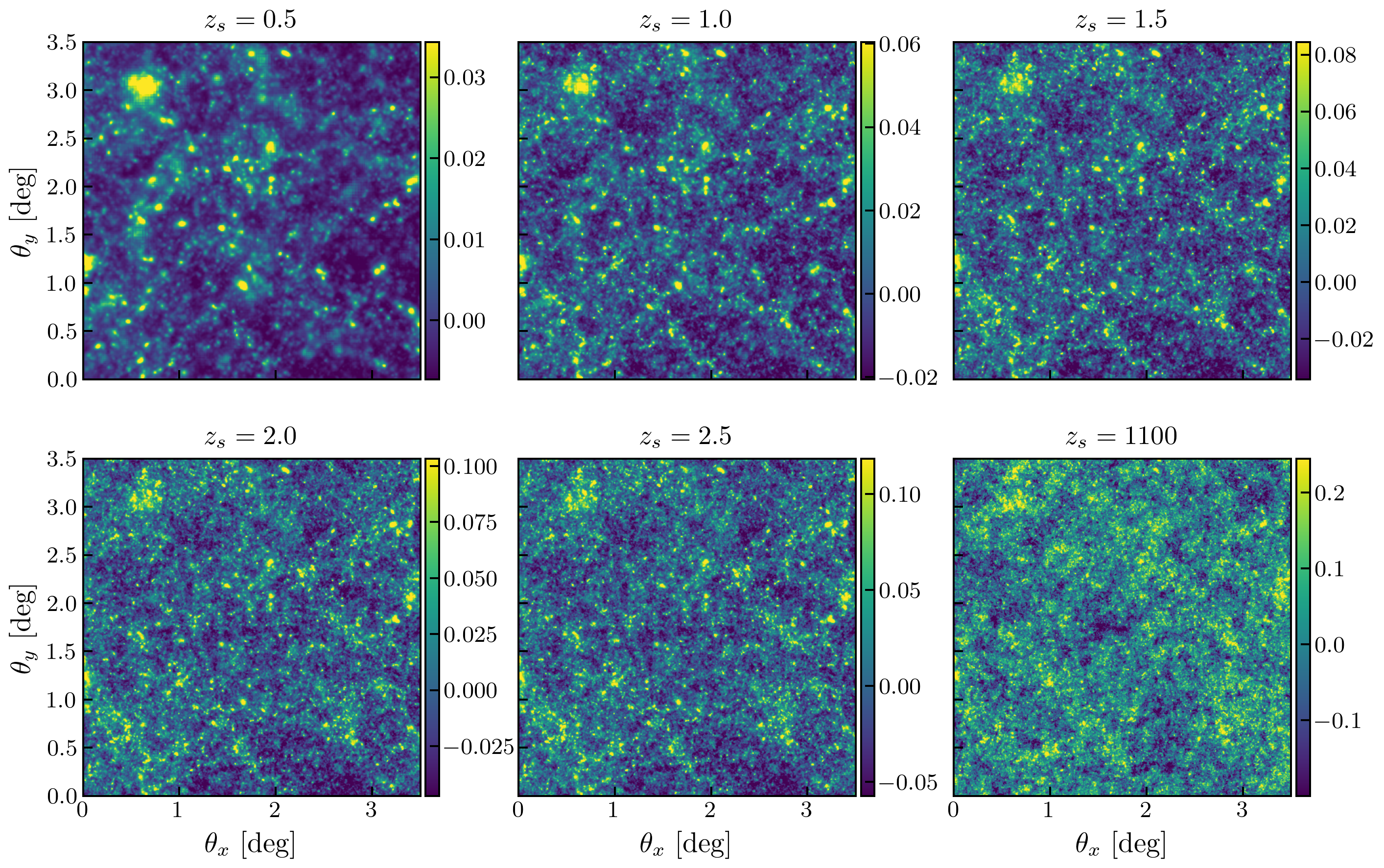}
    \caption{Lensing convergence maps $\kappa$ at six source redshifts $z_s \in \{0.5, 1.0, 1.5, 2.0, 2.5, 1100.0\}$ for the fiducial $\Lambda$CDM cosmology from a fixed initial condition.}
    \label{fig:maps_zs}
\end{figure*}

\begin{figure*}[t]
    \centering
    \includegraphics[width=\textwidth]{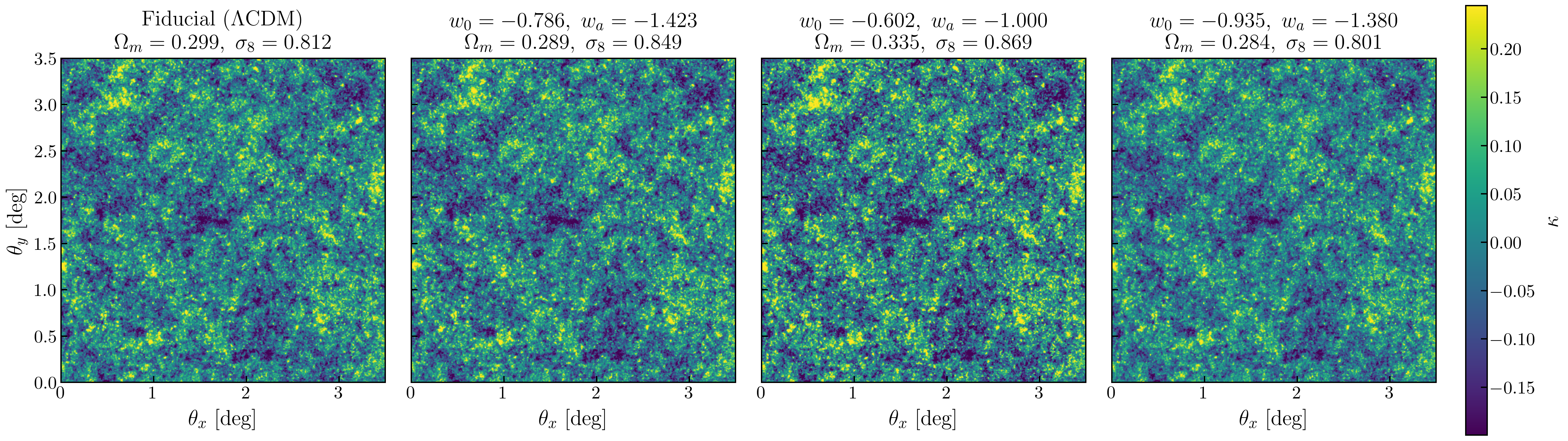}
    \caption{CMB lensing convergence maps $\kappa$ for four different cosmologies. This demonstrates how the structures depend on the cosmology with a fixed set of initial condition.}
    \label{fig:maps_cosmo}
\end{figure*}

We generate 10,000 realizations for each of our 128 dark energy cosmologies, and 100,000 realizations for the fiducial $\Lambda$CDM cosmology with three different ICs for the covariance matrix later on. Maps are generated at five source redshifts $z_s \in \{0.5, 1.0, 1.5, 2.0, 2.5\}$ as well as the CMB last-scattering surface at $z_s = 1100$\footnote{Since there is a redshift gap between $z=49$ and $z=1100$, we add an additional dummy plane at $z=1100+\epsilon$ when we run \texttt{LensTools}.}. In addition, as the number of snapshots varies when we get to $z=49$ for different cosmologies, we use the smallest number of snapshots out of all the cases we have to ensure the random seeds are correctly aligned among all the cases. Figure~\ref{fig:maps_zs} shows example convergence maps $\kappa$ at six source redshifts for the fiducial cosmology, illustrating how the lensing signal increases with source redshift as photons traverse more large-scale structure along the line of sight. We can also see that for fields with lower source redshifts, they are smoother because of the fixed resolution of our lens planes. We also show how CMB lensing convergence maps change over different sampled cosmologies in Figure~\ref{fig:maps_cosmo}. 


\subsection{Comparison with HMCode}
\begin{figure*}[t]
    \centering
    \includegraphics[width=\textwidth]{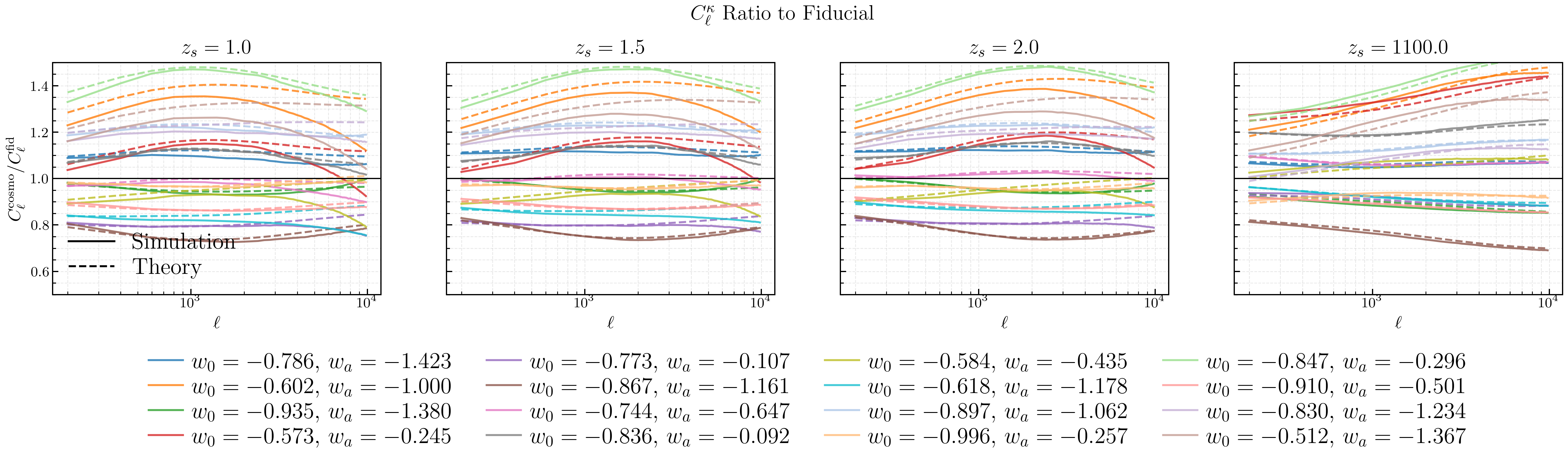}
    \caption{Ratio of the weak lensing convergence power spectrum $C_\ell^\kappa$ to the fiducial $\Lambda$CDM cosmology, for 16 randomly selected $w_0$--$w_a$ dark energy cosmologies at four source redshifts $z_s \in \{1.0, 1.5, 2.0, 1100\}$. Solid lines show simulation measurements and dashed lines show the corresponding HMcode-2020 predictions computed with \texttt{PYCCL}.}
    \label{fig:cl_ratio}
\end{figure*}

As a validation of our ray-tracing pipeline, we compare the measured weak lensing convergence power spectrum $C_\ell^\kappa$ against the theoretical prediction of HMcode-2020 computed with \texttt{PYCCL}~\cite{LSSTDarkEnergyScience:2018yem}\footnote{Code and documentation can be found in \url{https://github.com/LSSTDESC/CCL} and \url{https://ccl.readthedocs.io/en/latest/index.html}, respectively.} across our suite of dark energy cosmologies. Figure~\ref{fig:cl_ratio} shows the ratio of $C_\ell^\kappa$ to the fiducial $\Lambda$CDM cosmology for 16 randomly selected $w_0$--$w_a$ cosmologies at four source redshifts $z_s \in \{1.0, 1.5, 2.0, 1100\}$. Overall, we find good agreement between the simulation measurements and the HMcode-2020 predictions across all source redshifts, with deviations generally within the $\pm 5\%$ band on quasi-linear scales except for ones that are far away from $\Lambda$CDM. 

\subsection{Cross-correlation between weak lensing maps}
\begin{figure*}[t]
    \centering
    \includegraphics[width=\textwidth]{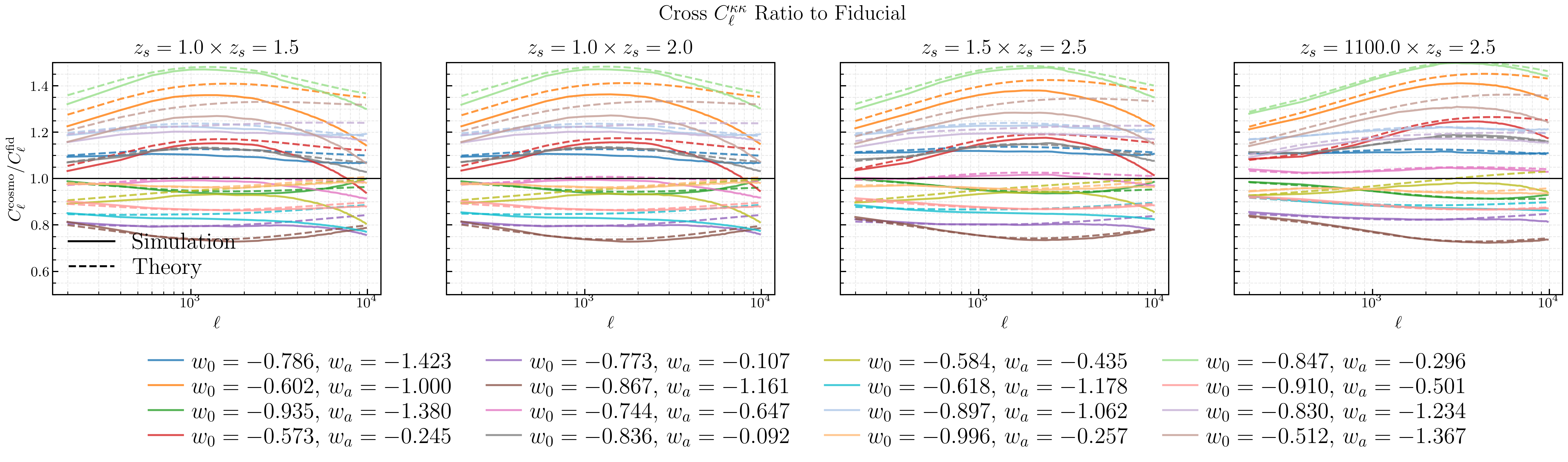}
    \caption{Ratio of the weak lensing convergence cross-power spectrum $C_\ell^{\kappa_a \kappa_b}$ to the fiducial $\Lambda$CDM cosmology, for 16 randomly selected $w_0$--$w_a$ dark energy cosmologies at four source redshift pairs $(z_a, z_b) \in \{(1.0, 1.5),\,(1.0, 2.0),\,(1.5, 2.5),\,(2.5, 1100)\}$. Solid lines show simulation measurements and dashed lines show the corresponding HMcode-2020 predictions computed with \texttt{PYCCL} and \texttt{CAMB}.}
    \label{fig:cl_cross_ratio}
\end{figure*}

When generating weak lensing maps with different source redshifts, we use the same random seed to generate the lens planes starting from $z=0$. This ensures that the line-of-sight structure is shared between maps at different source redshifts, so that maps are correlated with one another. Figure~\ref{fig:cl_cross_ratio} shows the ratio of the cross-spectrum $C_\ell^{\kappa_a \kappa_b}$ to the fiducial $\Lambda$CDM cosmology for 16 randomly selected $w_0$--$w_a$ dark energy cosmologies at four source redshift pairs $(z_a, z_b) \in \{(1.0, 1.5),\,(1.0, 2.0),\,(1.5, 2.5),\,(2.5, 1100)\}$, demonstrating that the cross-spectra carry complementary cosmological information to the auto-spectra and are well reproduced by the HMcode-2020 predictions computed with \texttt{PYCCL} across the full range of scales considered.

\subsection{Cross-correlation between weak lensing maps and halo overdensity fields}
Cross-correlation between the halo overdensity field and weak lensing maps can also be performed with our simulation suite. To achieve this, we use the halo catalogue obtained in Section~\ref{sec:haloCatalogue} and apply the same lens plane generation method to project the halo catalogue onto a two-dimensional plane, obtaining the projected number overdensity fields for each snapshot.\footnote{We modify the \texttt{LensTools} code to generate the number overdensity field alongside the original mass density field.} We then extract the lens plane selection and random shifts used during ray-tracing to align the halo overdensity field consistently with each weak lensing map. 

\begin{figure*}[t]
    \centering
    \includegraphics[width=0.335\textwidth]{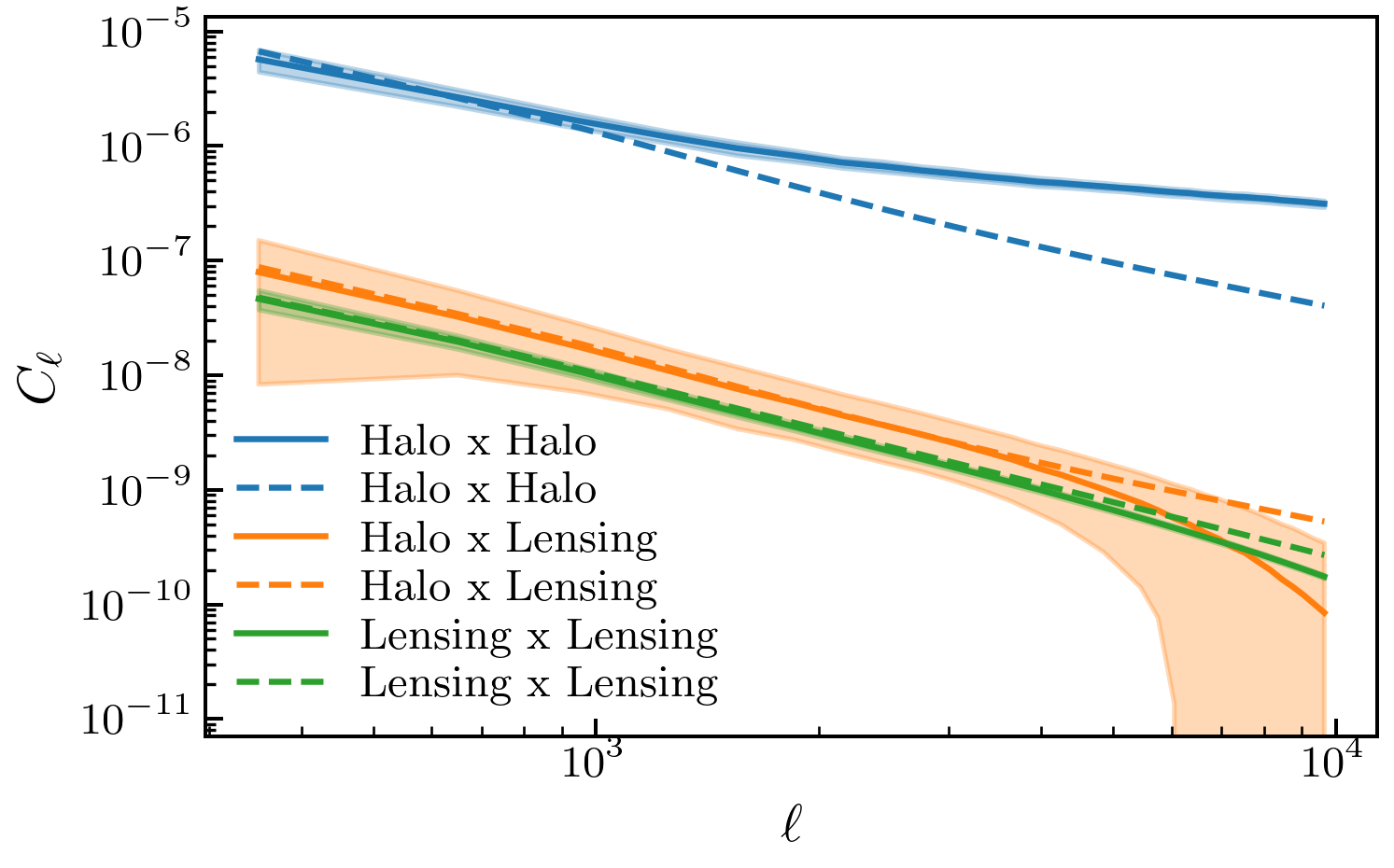}
    \includegraphics[width=0.32\textwidth]{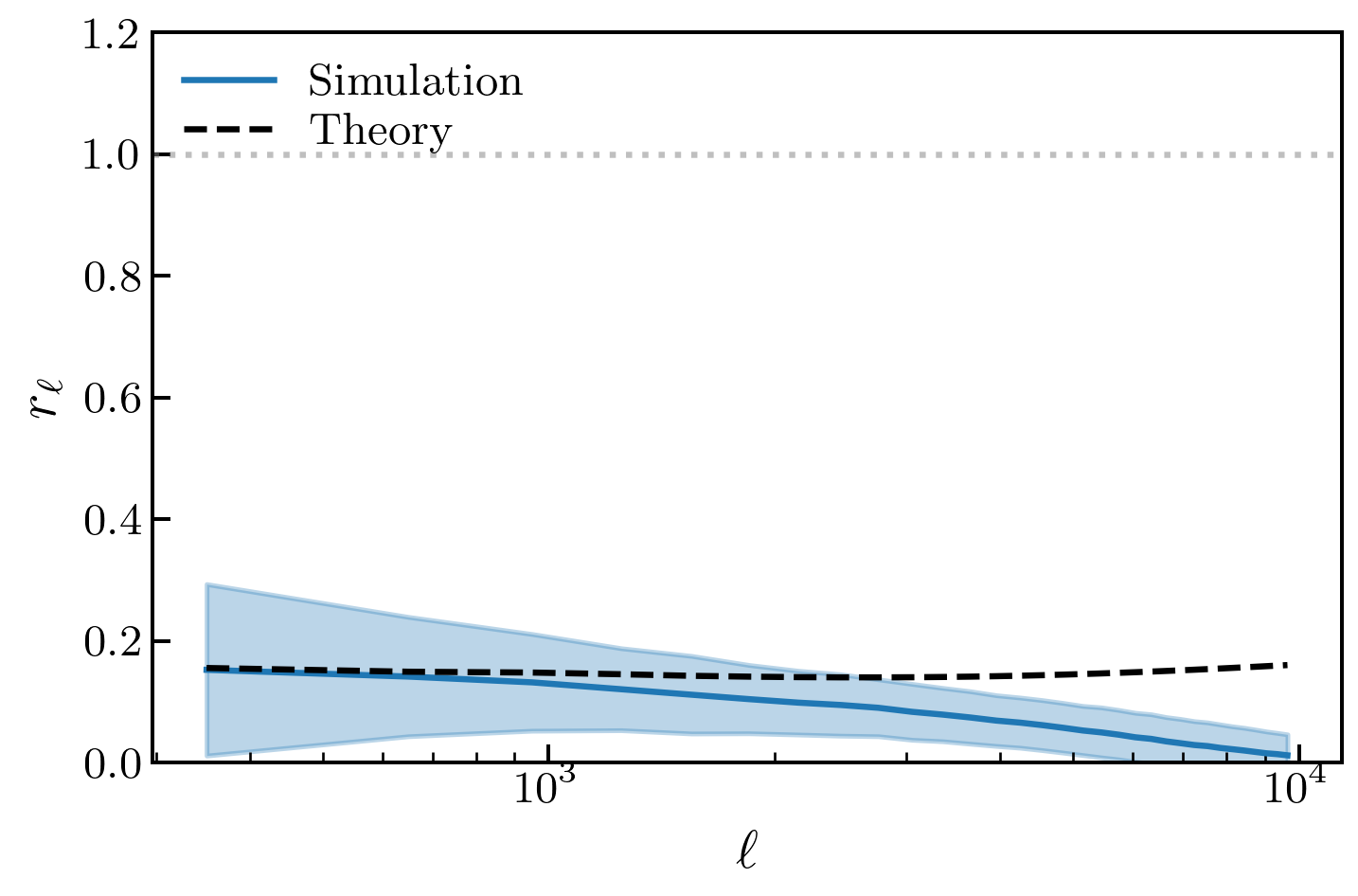}
    \caption{Angular power spectra (Left) and correlation coefficient (Right) between the projected halo density field centered at $z = 2.2$ and CMB lensing convergence $\kappa$, from 10,000 realizations of the fiducial cosmology. Halos with more than 100 particles are all included. Shaded bands show $1\sigma$ scatter. Dashed lines show linear theory with bias $b_1 = 3.46$ fitted on quasi-linear scales ($100 \leq \ell \leq 1000$).}
    \label{fig:halo_x_cmblensing}
\end{figure*}

Figure~\ref{fig:halo_x_cmblensing} shows the resulting angular power spectra and correlation coefficient between the projected halo overdensity field at $z = 2.2$ and the CMB lensing convergence $\kappa$, measured from 10,000 realizations of the fiducial cosmology using all halos with more than 100 particles and a top-hat window function of thickness $\pm 90\,\mathrm{Mpc}$ in comoving distance centered at the snapshot redshift. Shaded bands indicate the $1\sigma$ scatter across realizations. Dashed lines show the linear theory prediction with a scale-independent bias $b_1 = 3.46$ fitted on quasi-linear scales ($100 \leq \ell \leq 1000$).  The correlation coefficient $r_\ell = C_\ell^{h\kappa} / \sqrt{C_\ell^{hh} C_\ell^{\kappa\kappa}}$ is nonzero, confirming that the halo field coherently traces the underlying matter distribution.

\section{Conclusions}\label{sec:conclusion}
We have presented a suite of 128 N-body simulations and ray-traced weak lensing convergence maps in $w_0w_a$CDM cosmologies with massive neutrinos, designed to support simulation-based inference of dynamical dark energy parameters from weak gravitational lensing observables. Our pipeline combines several components: nonlinear evolution with \texttt{PKDGRAV3} incorporating relativistic corrections from all linear species, halo finding with \texttt{ROCKSTAR}, and fully ray-traced convergence maps generated with \texttt{LensTools}.

For each of the 128 cosmologies, sampled with a Sobol sequence over $\Omega_m \in [0.28, 0.35]$, $\sigma_8 \in [0.74, 0.90]$, $w_0 \in [-1.0, -0.5]$, and $w_a \in [-1.5, 0.0]$, we generate 10,000 ray-traced convergence maps at six source redshifts $z_s \in \{0.5, 1.0, 1.5, 2.0, 2.5, 1100\}$, and an additional 100,000 realizations for the fiducial $\Lambda$CDM cosmology to enable robust covariance estimation. We additionally provide halo catalogues with three mass definitions ($M_\mathrm{vir}$, $M_\mathrm{200b}$, $M_\mathrm{200c}$) and projected halo overdensity fields, enabling galaxy--lensing cross-correlation analyses. We have validated our pipeline at each stage against theoretical predictions: the matter power spectrum and lensing convergence power spectrum agree with HMcode-2020, and the halo mass function agrees with the Tinker et al.~(2008) fitting function across all three mass definitions and all sampled cosmologies.

This simulation suite opens up several avenues for future work. The most immediate application is simulation-based inference of $w_0$ and $w_a$ using non-Gaussian lensing statistics such as peak counts~\cite{Bardeen:1985tr,1987MNRAS.226..655B}, Minkowski functionals~\cite{Minkowski1903}, wavelet scattering coefficients~\cite{2011arXiv1101.2286M}, or field-level inference with neural posterior estimators — which can substantially tighten constraints beyond what is achievable with the power spectrum alone. The availability of maps at multiple source redshifts enables tomographic analyses and cross-correlation studies, including cross-correlation with CMB lensing from experiments such as the Simons Observatory~\cite{SimonsObservatory:2018koc,SimonsObservatory:2025wwn}. The halo catalogues further enable studies of galaxy--lensing cross-correlations and halo model validation in $w_0w_a$CDM cosmologies. Looking further ahead, extending the suite to include baryonic effects, intrinsic alignments, and a broader range of dark energy models would broaden its applicability to the full science programs of Stage-IV surveys such as Rubin LSST~\cite{LSSTDarkEnergyScience:2018yem}, Euclid~\cite{Amendola:2016saw} and the Roman Space Telescope~\cite{Akeson:2019biv,Spergel:2015sza}. 

\section{Acknowledgments}
SFC, JCH, and ZH acknowledge support from NASA grant 80NSSC24K1093 [ATP]. The authors also acknowledge the Texas Advanced Computing Center (TACC) at The University of Texas at Austin for providing computational resources that have contributed to the research results reported within this paper. We acknowledge computing resources from Columbia University’s Shared Research Computing Facility project, which is supported by NIH Research Facility Improvement Grant 1G20RR030893-01, and associated funds from the New York State Empire State Development, Division of Science Technology and Innovation (NYSTAR) Contract C090171, both awarded April 15, 2010. We will make the simulation suite publicly available upon acceptance of the draft.

\clearpage
\newpage

\bibliography{ref.bib}

@article{DESI:2024mwx,
    author = "Adame, A. G. and others",
    collaboration = "DESI",
    title = "{DESI 2024 VI: cosmological constraints from the measurements of baryon acoustic oscillations}",
    eprint = "2404.03002",
    archivePrefix = "arXiv",
    primaryClass = "astro-ph.CO",
    reportNumber = "FERMILAB-PUB-24-0154-PPD",
    doi = "10.1088/1475-7516/2025/02/021",
    journal = "JCAP",
    volume = "02",
    pages = "021",
    year = "2025"
}

@article{DESI:2024jxi,
    author = "Adame, A. G. and others",
    collaboration = "DESI",
    title = "{DESI 2024 V: Full-Shape galaxy clustering from galaxies and quasars}",
    eprint = "2411.12021",
    archivePrefix = "arXiv",
    primaryClass = "astro-ph.CO",
    reportNumber = "FERMILAB-PUB-24-0847-PPD",
    doi = "10.1088/1475-7516/2025/09/008",
    journal = "JCAP",
    volume = "09",
    pages = "008",
    year = "2025",
    note = "[Erratum: JCAP 02, E02 (2026)]"
}

@article{DESI:2024hhd,
    author = "Adame, A. G. and others",
    collaboration = "DESI",
    title = "{DESI 2024 VII: cosmological constraints from the full-shape modeling of clustering measurements}",
    eprint = "2411.12022",
    archivePrefix = "arXiv",
    primaryClass = "astro-ph.CO",
    reportNumber = "FERMILAB-PUB-24-0854-PPD",
    doi = "10.1088/1475-7516/2025/07/028",
    journal = "JCAP",
    volume = "07",
    pages = "028",
    year = "2025"
}

@article{DESI:2025zgx,
    author = "Abdul Karim, M. and others",
    collaboration = "DESI",
    title = "{DESI DR2 results. II. Measurements of baryon acoustic oscillations and cosmological constraints}",
    eprint = "2503.14738",
    archivePrefix = "arXiv",
    primaryClass = "astro-ph.CO",
    reportNumber = "FERMILAB-PUB-25-0169-PPD",
    doi = "10.1103/tr6y-kpc6",
    journal = "Phys. Rev. D",
    volume = "112",
    number = "8",
    pages = "083515",
    year = "2025"
}

@article{DESI:2025zpo,
    author = "Abdul Karim, M. and others",
    collaboration = "DESI",
    title = "{DESI DR2 results. I. Baryon acoustic oscillations from the Lyman alpha forest}",
    eprint = "2503.14739",
    archivePrefix = "arXiv",
    primaryClass = "astro-ph.CO",
    reportNumber = "FERMILAB-PUB-25-0167-PPD",
    doi = "10.1103/2wwn-xjm5",
    journal = "Phys. Rev. D",
    volume = "112",
    number = "8",
    pages = "083514",
    year = "2025"
}

@article{Brout:2022vxf,
    author = "Brout, Dillon and others",
    title = "{The Pantheon+ Analysis: Cosmological Constraints}",
    eprint = "2202.04077",
    archivePrefix = "arXiv",
    primaryClass = "astro-ph.CO",
    doi = "10.3847/1538-4357/ac8e04",
    journal = "Astrophys. J.",
    volume = "938",
    number = "2",
    pages = "110",
    year = "2022"
}

@article{DES:2024jxu,
    author = "Abbott, T. M. C. and others",
    collaboration = "DES",
    title = "{The Dark Energy Survey: Cosmology Results with {\ensuremath{\sim}}1500 New High-redshift Type Ia Supernovae Using the Full 5 yr Data Set}",
    eprint = "2401.02929",
    archivePrefix = "arXiv",
    primaryClass = "astro-ph.CO",
    reportNumber = "FERMILAB-PUB-23-0821-PPD, DES-2023-805",
    doi = "10.3847/2041-8213/ad6f9f",
    journal = "Astrophys. J. Lett.",
    volume = "973",
    number = "1",
    pages = "L14",
    year = "2024"
}

@article{Rubin:2023jdq,
    author = "Rubin, David and others",
    title = "{Union Through UNITY: Cosmology with 2,000 SNe Using a Unified Bayesian Framework}",
    eprint = "2311.12098",
    archivePrefix = "arXiv",
    primaryClass = "astro-ph.CO",
    doi = "10.3847/1538-4357/adc0a5",
    journal = "Astrophys. J.",
    volume = "986",
    number = "2",
    pages = "231",
    year = "2025"
}

@article{DES:2025sig,
    author = "Popovic, B. and others",
    collaboration = "DES",
    title = "{The Dark Energy Survey Supernova Program: A Reanalysis Of Cosmology Results And Evidence For Evolving Dark Energy With An Updated Type Ia Supernova Calibration}",
    eprint = "2511.07517",
    archivePrefix = "arXiv",
    primaryClass = "astro-ph.CO",
    reportNumber = "FERMILAB-PUB-25-0842-CSAID-PPD",
    doi = "10.1093/mnras/stag632",
    month = "11",
    year = "2025"
}

@article{Gialamas:2024lyw,
    author = {Gialamas, Ioannis D. and H{\"u}tsi, Gert and Kannike, Kristjan and Racioppi, Antonio and Raidal, Martti and Vasar, Martin and Veerm{\"a}e, Hardi},
    title = "{Interpreting DESI 2024 BAO: Late-time dynamical dark energy or a local effect?}",
    eprint = "2406.07533",
    archivePrefix = "arXiv",
    primaryClass = "astro-ph.CO",
    doi = "10.1103/PhysRevD.111.043540",
    journal = "Phys. Rev. D",
    volume = "111",
    number = "4",
    pages = "043540",
    year = "2025"
}

@article{Dhawan:2024gqy,
    author = "Dhawan, Suhail and Popovic, Brodie and Goobar, Ariel",
    title = "{The axis of systematic bias in SN Ia cosmology and implications for DESI 2024 results}",
    eprint = "2409.18668",
    archivePrefix = "arXiv",
    primaryClass = "astro-ph.CO",
    doi = "10.1093/mnras/staf779",
    journal = "Mon. Not. Roy. Astron. Soc.",
    volume = "540",
    number = "2",
    pages = "1626--1634",
    year = "2025"
}

@article{Chen:2024vuf,
    author = "Chen, Shi-Fan and Ivanov, Mikhail M. and Philcox, Oliver H. E. and Wenzl, Lukas",
    title = "{Suppression without Thawing: Constraining Structure Formation and Dark Energy with Galaxy Clustering}",
    eprint = "2406.13388",
    archivePrefix = "arXiv",
    primaryClass = "astro-ph.CO",
    doi = "10.1103/PhysRevLett.133.231001",
    journal = "Phys. Rev. Lett.",
    volume = "133",
    number = "23",
    pages = "231001",
    year = "2024"
}

@article{Chen:2025jnr,
    author = "Chen, Shu-Fan and Ivanov, Mikhail M.",
    title = "{Constraining dynamical dark energy from galaxy clustering with simulation-based priors}",
    eprint = "2507.00118",
    archivePrefix = "arXiv",
    primaryClass = "astro-ph.CO",
    reportNumber = "MIT-CTP/5881",
    doi = "10.1103/3nbk-l5v1",
    journal = "Phys. Rev. D",
    volume = "112",
    number = "8",
    pages = "083562",
    year = "2025"
}

@article{Silva:2025twg,
    author = "Silva, Emanuelly and Nunes, Rafael C.",
    title = "{Testing signatures of phantom crossing through full-shape galaxy clustering analysis}",
    eprint = "2507.13989",
    archivePrefix = "arXiv",
    primaryClass = "astro-ph.CO",
    doi = "10.1088/1475-7516/2025/11/078",
    journal = "JCAP",
    volume = "11",
    pages = "078",
    year = "2025"
}

@article{Euclid:2020rfv,
    author = "Knabenhans, M. and others",
    collaboration = "Euclid",
    title = "{Euclid preparation: IX. EuclidEmulator2 {\textendash} power spectrum emulation with massive neutrinos and self-consistent dark energy perturbations}",
    eprint = "2010.11288",
    archivePrefix = "arXiv",
    primaryClass = "astro-ph.CO",
    doi = "10.1093/mnras/stab1366",
    journal = "Mon. Not. Roy. Astron. Soc.",
    volume = "505",
    number = "2",
    pages = "2840--2869",
    year = "2021"
}

@article{Angulo:2020vky,
    author = {Angulo, Raul E. and Zennaro, Matteo and Contreras, Sergio and Aric{\`o}, Giovanni and Pellejero-Iba{\~n}ez, Marcos and St{\"u}cker, Jens},
    title = "{The BACCO simulation project: exploiting the full power of large-scale structure for cosmology}",
    eprint = "2004.06245",
    archivePrefix = "arXiv",
    primaryClass = "astro-ph.CO",
    doi = "10.1093/mnras/stab2018",
    journal = "Mon. Not. Roy. Astron. Soc.",
    volume = "507",
    number = "4",
    pages = "5869--5881",
    year = "2021"
}

@article{Lawrence:2017ost,
    author = "Lawrence, Earl and Heitmann, Katrin and Kwan, Juliana and Upadhye, Amol and Bingham, Derek and Habib, Salman and Higdon, David and Pope, Adrian and Finkel, Hal and Frontiere, Nicholas",
    title = "{The Mira-Titan Universe II: Matter Power Spectrum Emulation}",
    eprint = "1705.03388",
    archivePrefix = "arXiv",
    primaryClass = "astro-ph.CO",
    doi = "10.3847/1538-4357/aa86a9",
    journal = "Astrophys. J.",
    volume = "847",
    number = "1",
    pages = "50",
    year = "2017"
}

@article{Chen:2025ugn,
    author = "Chen, Zhao and Yu, Yu and Han, Jiaxin and Jing, Yipeng",
    title = "{CSST cosmological emulator I: Matter power spectrum emulation with one percent accuracy to k = 10h Mpc$^{−1}$}",
    eprint = "2502.11160",
    archivePrefix = "arXiv",
    primaryClass = "astro-ph.CO",
    doi = "10.1007/s11433-025-2671-0",
    journal = "Sci. China Phys. Mech. Astron.",
    volume = "68",
    number = "8",
    pages = "289512",
    year = "2025"
}

@article{Zhou:2025iiu,
    author = "Zhou, Shuren and Chen, Zhao and Yu, Yu",
    title = "{CSST cosmological emulator III: Hybrid lagrangian bias expansion emulation of galaxy clustering}",
    eprint = "2506.04671",
    archivePrefix = "arXiv",
    primaryClass = "astro-ph.CO",
    doi = "10.1007/s11433-025-2755-x",
    journal = "Sci. China Phys. Mech. Astron.",
    volume = "68",
    number = "12",
    pages = "129512",
    year = "2025"
}

@article{Chen:2025opy,
    author = "Chen, Zhao and Yu, Yu",
    title = "{CSST cosmological emulator II: Generalized accurate halo mass function emulation}",
    eprint = "2506.09688",
    archivePrefix = "arXiv",
    primaryClass = "astro-ph.CO",
    reportNumber = "Science China Physics, Mechanics {\&} Astronomy, Volume 68, Issue 10, 2025, 109513",
    doi = "10.1007/s11433-025-2764-x",
    journal = "Sci. China Phys. Mech. Astron.",
    volume = "68",
    number = "10",
    pages = "109513",
    year = "2025"
}

@article{DeRose:2023dmk,
    author = "DeRose, Joseph and Kokron, Nickolas and Banerjee, Arka and Chen, Shi-Fan and White, Martin and Wechsler, Risa and Storey-Fisher, Kate and Tinker, Jeremy and Zhai, Zhongxu",
    title = "{Aemulus {\ensuremath{\nu}}: precise predictions for matter and biased tracer power spectra in the presence of neutrinos}",
    eprint = "2303.09762",
    archivePrefix = "arXiv",
    primaryClass = "astro-ph.CO",
    doi = "10.1088/1475-7516/2023/07/054",
    journal = "JCAP",
    volume = "07",
    pages = "054",
    year = "2023"
}

@article{DeRose:2018xdj,
    author = "DeRose, Joseph and Wechsler, Risa H. and Tinker, Jeremy L. and Becker, Matthew R. and Mao, Yao-Yuan and McClintock, Thomas and McLaughlin, Sean and Rozo, Eduardo and Zhai, Zhongxu",
    title = "{The Aemulus Project I: Numerical Simulations for Precision Cosmology}",
    eprint = "1804.05865",
    archivePrefix = "arXiv",
    primaryClass = "astro-ph.CO",
    doi = "10.3847/1538-4357/ab1085",
    journal = "Astrophys. J.",
    volume = "875",
    number = "1",
    pages = "69",
    year = "2019"
}

@article{McClintock:2018uyf,
    author = "McClintock, Thomas and Rozo, Eduardo and Becker, Matthew R. and DeRose, Joseph and Mao, Yau-Yuan and McLaughlin, Sean and Tinker, Jeremy L. and Wechsler, Risa H. and Zhai, Zhongxu",
    title = "{The Aemulus Project II: Emulating the Halo Mass Function}",
    eprint = "1804.05866",
    archivePrefix = "arXiv",
    primaryClass = "astro-ph.CO",
    doi = "10.3847/1538-4357/aaf568",
    journal = "Astrophys. J.",
    volume = "872",
    number = "1",
    pages = "53",
    year = "2019"
}

@article{Zhai:2018plk,
    author = "Zhai, Zhongxu and Tinker, Jeremy L. and Becker, Matthew R. and DeRose, Joseph and Mao, Yao-Yuan and McClintock, Thomas and McLaughlin, Sean and Rozo, Eduardo and Wechsler, Risa H.",
    title = "{The Aemulus Project III: Emulation of the Galaxy Correlation Function}",
    eprint = "1804.05867",
    archivePrefix = "arXiv",
    primaryClass = "astro-ph.CO",
    doi = "10.3847/1538-4357/ab0d7b",
    journal = "Astrophys. J.",
    volume = "874",
    number = "1",
    pages = "95",
    year = "2019"
}

@article{Yang:2025dtc,
    author = "Yang, Yanhui and Bird, Simeon and Ho, Ming-Feng",
    title = "{Ten-parameter simulation suite for cosmological emulation beyond {\ensuremath{\Lambda}}CDM}",
    eprint = "2501.06296",
    archivePrefix = "arXiv",
    primaryClass = "astro-ph.CO",
    doi = "10.1103/PhysRevD.111.083529",
    journal = "Phys. Rev. D",
    volume = "111",
    number = "8",
    pages = "083529",
    year = "2025"
}

@article{Blot:2026var,
    author = "Blot, Linda and Gayoux, Th{\'e}o and Schmidt, Fabian and Corasaniti, Pier Stefano and de Ligondes, Bastien",
    title = "{Cosmological hydrodynamical simulations of clustering dark energy with Nefertiti}",
    eprint = "2603.16248",
    archivePrefix = "arXiv",
    primaryClass = "astro-ph.CO",
    month = "3",
    year = "2026"
}

@article{Blot:2022mnt,
    author = "Blot, Linda and Corasaniti, Pier Stefano and Schmidt, Fabian",
    title = "{Non-linear Eulerian hydrodynamics of dark energy: Riemann problem and finite volume schemes}",
    eprint = "2210.04800",
    archivePrefix = "arXiv",
    primaryClass = "astro-ph.CO",
    doi = "10.1088/1475-7516/2023/05/001",
    journal = "JCAP",
    volume = "05",
    pages = "001",
    year = "2023"
}

@article{Creminelli:2009mu,
    author = "Creminelli, Paolo and D'Amico, Guido and Norena, Jorge and Senatore, Leonardo and Vernizzi, Filippo",
    title = "{Spherical collapse in quintessence models with zero speed of sound}",
    eprint = "0911.2701",
    archivePrefix = "arXiv",
    primaryClass = "astro-ph.CO",
    doi = "10.1088/1475-7516/2010/03/027",
    journal = "JCAP",
    volume = "03",
    pages = "027",
    year = "2010"
}

@article{Sefusatti:2011cm,
    author = "Sefusatti, Emiliano and Vernizzi, Filippo",
    title = "{Cosmological structure formation with clustering quintessence}",
    eprint = "1101.1026",
    archivePrefix = "arXiv",
    primaryClass = "astro-ph.CO",
    reportNumber = "T11-007",
    doi = "10.1088/1475-7516/2011/03/047",
    journal = "JCAP",
    volume = "03",
    pages = "047",
    year = "2011"
}

@article{Reboucas:2025hue,
    author = "Rebou{\c{c}}as, Jo{\~a}o and Lloyd, Victoria and Gordon, Jonathan and Brando, Guilherme and Miranda, Vivian",
    title = "{Modeling nonlinear scales for dynamical dark energy cosmologies with COLA}",
    eprint = "2510.14888",
    archivePrefix = "arXiv",
    primaryClass = "astro-ph.CO",
    doi = "10.1103/1zgv-sx5s",
    journal = "Phys. Rev. D",
    volume = "113",
    number = "10",
    pages = "103513",
    year = "2026"
}

@article{Springel:2005mi,
    author = "Springel, Volker",
    title = "{The Cosmological simulation code GADGET-2}",
    eprint = "astro-ph/0505010",
    archivePrefix = "arXiv",
    doi = "10.1111/j.1365-2966.2005.09655.x",
    journal = "Mon. Not. Roy. Astron. Soc.",
    volume = "364",
    pages = "1105--1134",
    year = "2005"
}

@article{Potter:2016ttn,
    author = "Potter, Douglas and Stadel, Joachim and Teyssier, Romain",
    title = "{PKDGRAV3: beyond trillion particle cosmological simulations for the next era of galaxy surveys}",
    eprint = "1609.08621",
    archivePrefix = "arXiv",
    primaryClass = "astro-ph.IM",
    doi = "10.1186/s40668-017-0021-1",
    journal = "Comput. Astrophys. Cosmol.",
    volume = "4",
    number = "1",
    pages = "2",
    year = "2017"
}

@ARTICLE{1987JCoPh..73..325G,
       author = {{Greengard}, L. and {Rokhlin}, V.},
        title = "{A Fast Algorithm for Particle Simulations}",
      journal = {Journal of Computational Physics},
         year = 1987,
        month = dec,
       volume = {73},
       number = {2},
        pages = {325-348},
          doi = {10.1016/0021-9991(87)90140-9},
       adsurl = {https://ui.adsabs.harvard.edu/abs/1987JCoPh..73..325G}
}

@ARTICLE{2002JCoPh.179...27D,
       author = {{Dehnen}, Walter},
        title = "{A Hierarchical <E10>O</E10>(N) Force Calculation Algorithm}",
      journal = {Journal of Computational Physics},
         year = 2002,
        month = jun,
       volume = {179},
       number = {1},
        pages = {27-42},
          doi = {10.1006/jcph.2002.7026},
archivePrefix = {arXiv},
       eprint = {astro-ph/0202512},
 primaryClass = {astro-ph},
       adsurl = {https://ui.adsabs.harvard.edu/abs/2002JCoPh.179...27D}
}

@article{Dakin:2019dxu,
    author = "Dakin, Jeppe and Hannestad, Steen and Tram, Thomas",
    title = "{Fully relativistic treatment of decaying cold dark matter in $N$-body simulations}",
    eprint = "1904.11773",
    archivePrefix = "arXiv",
    primaryClass = "astro-ph.CO",
    doi = "10.1088/1475-7516/2019/06/032",
    journal = "JCAP",
    volume = "06",
    pages = "032",
    year = "2019"
}

@article{Dakin:2019vnj,
    author = "Dakin, Jeppe and Hannestad, Steen and Tram, Thomas and Knabenhans, Mischa and Stadel, Joachim",
    title = "{Dark energy perturbations in $N$-body simulations}",
    eprint = "1904.05210",
    archivePrefix = "arXiv",
    primaryClass = "astro-ph.CO",
    doi = "10.1088/1475-7516/2019/08/013",
    journal = "JCAP",
    volume = "08",
    pages = "013",
    year = "2019"
}

@article{Chevallier:2000qy,
    author = "Chevallier, Michel and Polarski, David",
    title = "{Accelerating universes with scaling dark matter}",
    eprint = "gr-qc/0009008",
    archivePrefix = "arXiv",
    doi = "10.1142/S0218271801000822",
    journal = "Int. J. Mod. Phys. D",
    volume = "10",
    pages = "213--224",
    year = "2001"
}

@article{Linder:2002et,
    author = "Linder, Eric V.",
    title = "{Exploring the expansion history of the universe}",
    eprint = "astro-ph/0208512",
    archivePrefix = "arXiv",
    doi = "10.1103/PhysRevLett.90.091301",
    journal = "Phys. Rev. Lett.",
    volume = "90",
    pages = "091301",
    year = "2003"
}

@article{Baker:2012zs,
    author = "Baker, Tessa and Ferreira, Pedro G. and Skordis, Constantinos",
    title = "{The Parameterized Post-Friedmann framework for theories of modified gravity: concepts, formalism and examples}",
    eprint = "1209.2117",
    archivePrefix = "arXiv",
    primaryClass = "astro-ph.CO",
    doi = "10.1103/PhysRevD.87.024015",
    journal = "Phys. Rev. D",
    volume = "87",
    number = "2",
    pages = "024015",
    year = "2013"
}

@article{Fidler:2016tir,
    author = "Fidler, Christian and Tram, Thomas and Rampf, Cornelius and Crittenden, Robert and Koyama, Kazuya and Wands, David",
    title = "{Relativistic Interpretation of Newtonian Simulations for Cosmic Structure Formation}",
    eprint = "1606.05588",
    archivePrefix = "arXiv",
    primaryClass = "astro-ph.CO",
    doi = "10.1088/1475-7516/2016/09/031",
    journal = "JCAP",
    volume = "09",
    pages = "031",
    year = "2016"
}

@article{Fidler:2017ebh,
    author = "Fidler, Christian and Tram, Thomas and Rampf, Cornelius and Crittenden, Robert and Koyama, Kazuya and Wands, David",
    title = "{Relativistic initial conditions for N-body simulations}",
    eprint = "1702.03221",
    archivePrefix = "arXiv",
    primaryClass = "astro-ph.CO",
    doi = "10.1088/1475-7516/2017/06/043",
    journal = "JCAP",
    volume = "06",
    pages = "043",
    year = "2017"
}

@article{Liu:2017now,
    author = "Liu, Jia and Bird, Simeon and Matilla, Jos{\'e} Manuel Zorrilla and Hill, J. Colin and Haiman, Zolt{\'a}n and Madhavacheril, Mathew S. and Petri, Andrea and Spergel, David N.",
    title = "{MassiveNuS: Cosmological Massive Neutrino Simulations}",
    eprint = "1711.10524",
    archivePrefix = "arXiv",
    primaryClass = "astro-ph.CO",
    doi = "10.1088/1475-7516/2018/03/049",
    journal = "JCAP",
    volume = "03",
    pages = "049",
    year = "2018"
}

@article{AtacamaCosmologyTelescope:2025blo,
    author = "Louis, Thibaut and others",
    collaboration = "Atacama Cosmology Telescope",
    title = "{The Atacama Cosmology Telescope: DR6 power spectra, likelihoods and {\ensuremath{\Lambda}}CDM parameters}",
    eprint = "2503.14452",
    archivePrefix = "arXiv",
    primaryClass = "astro-ph.CO",
    reportNumber = "FERMILAB-PUB-25-0071-PPD",
    doi = "10.1088/1475-7516/2025/11/062",
    journal = "JCAP",
    volume = "11",
    pages = "062",
    year = "2025"
}

@article{Sobol1967OnTD,
  title={On the distribution of points in a cube and the approximate evaluation of integrals},
  author={Ilya M. Sobol},
  journal={Ussr Computational Mathematics and Mathematical Physics},
  year={1967},
  volume={7},
  pages={86-112},
  url={https://api.semanticscholar.org/CorpusID:122581245}
}

@MISC{Pylians,
    author = {{Villaescusa-Navarro}, Francisco},
    title = "{Pylians: Python libraries for the analysis of numerical simulations}",
    howpublished = {Astrophysics Source Code Library, record ascl:1811.008},
    year = 2018,
    month = nov,
    eid = {ascl:1811.008},
    pages = {ascl:1811.008},
    archivePrefix = {ascl},
    eprint = {1811.008},
    adsurl = {https://ui.adsabs.harvard.edu/abs/2018ascl.soft11008V}
}

@article{Mead:2020vgs,
    author = {Mead, Alexander and Brieden, Samuel and Tr{\"o}ster, Tilman and Heymans, Catherine},
    title = "{hmcode-2020: improved modelling of non-linear cosmological power spectra with baryonic feedback}",
    eprint = "2009.01858",
    archivePrefix = "arXiv",
    primaryClass = "astro-ph.CO",
    doi = "10.1093/mnras/stab082",
    journal = "Mon. Not. Roy. Astron. Soc.",
    volume = "502",
    number = "1",
    pages = "1401--1422",
    year = "2021"
}

@article{Petri:2016zrq,
    author = "Petri, Andrea",
    title = "{Mocking the Weak Lensing universe: the LensTools python computing package}",
    eprint = "1606.01903",
    archivePrefix = "arXiv",
    primaryClass = "astro-ph.CO",
    doi = "10.1016/j.ascom.2016.06.001",
    journal = "Astron. Comput.",
    volume = "17",
    pages = "73--79",
    year = "2016"
}

@article{Blas:2011rf,
    author = "Blas, Diego and Lesgourgues, Julien and Tram, Thomas",
    title = "{The Cosmic Linear Anisotropy Solving System (CLASS) II: Approximation schemes}",
    eprint = "1104.2933",
    archivePrefix = "arXiv",
    primaryClass = "astro-ph.CO",
    reportNumber = "CERN-PH-TH-2011-082, LAPTH-010-11",
    doi = "10.1088/1475-7516/2011/07/034",
    journal = "JCAP",
    volume = "07",
    pages = "034",
    year = "2011"
}

@article{Behroozi:2011ju,
    author = "Behroozi, Peter S. and Wechsler, Risa H. and Wu, Hao-Yi",
    title = "{The Rockstar Phase-Space Temporal Halo Finder and the Velocity Offsets of Cluster Cores}",
    eprint = "1110.4372",
    archivePrefix = "arXiv",
    primaryClass = "astro-ph.CO",
    reportNumber = "SLAC-PUB-15331",
    doi = "10.1088/0004-637X/762/2/109",
    journal = "Astrophys. J.",
    volume = "762",
    pages = "109",
    year = "2013"
}

@article{Tinker:2008ff,
    author = "Tinker, Jeremy L. and Kravtsov, Andrey V. and Klypin, Anatoly and Abazajian, Kevork and Warren, Michael S. and Yepes, Gustavo and Gottlober, Stefan and Holz, Daniel E.",
    title = "{Toward a halo mass function for precision cosmology: The Limits of universality}",
    eprint = "0803.2706",
    archivePrefix = "arXiv",
    primaryClass = "astro-ph",
    doi = "10.1086/591439",
    journal = "Astrophys. J.",
    volume = "688",
    pages = "709--728",
    year = "2008"
}

@article{Diemer:2017bwl,
    author = "Diemer, Benedikt",
    title = "{COLOSSUS: A python toolkit for cosmology, large-scale structure, and dark matter halos}",
    eprint = "1712.04512",
    archivePrefix = "arXiv",
    primaryClass = "astro-ph.CO",
    doi = "10.3847/1538-4365/aaee8c",
    journal = "Astrophys. J. Suppl.",
    volume = "239",
    number = "2",
    pages = "35",
    year = "2018"
}

@article{Petri:2016wlu,
    author = "Petri, Andrea and Haiman, Zolt{\'a}n and May, Morgan",
    title = "{Sample variance in weak lensing: how many simulations are required?}",
    eprint = "1601.06792",
    archivePrefix = "arXiv",
    primaryClass = "astro-ph.CO",
    doi = "10.1103/PhysRevD.93.063524",
    journal = "Phys. Rev. D",
    volume = "93",
    number = "6",
    pages = "063524",
    year = "2016"
}

@article{LSSTDarkEnergyScience:2018yem,
    author = "Chisari, Nora Elisa and others",
    collaboration = "LSST Dark Energy Science",
    title = "{Core Cosmology Library: Precision Cosmological Predictions for LSST}",
    eprint = "1812.05995",
    archivePrefix = "arXiv",
    primaryClass = "astro-ph.CO",
    doi = "10.3847/1538-4365/ab1658",
    journal = "Astrophys. J. Suppl.",
    volume = "242",
    number = "1",
    pages = "2",
    year = "2019"
}

@article{SimonsObservatory:2018koc,
    author = "Ade, Peter and others",
    collaboration = "Simons Observatory",
    title = "{The Simons Observatory: Science goals and forecasts}",
    eprint = "1808.07445",
    archivePrefix = "arXiv",
    primaryClass = "astro-ph.CO",
    doi = "10.1088/1475-7516/2019/02/056",
    journal = "JCAP",
    volume = "02",
    pages = "056",
    year = "2019"
}

@article{SimonsObservatory:2025wwn,
    author = "Abitbol, M. and others",
    collaboration = "Simons Observatory",
    title = "{The Simons Observatory: science goals and forecasts for the enhanced Large Aperture Telescope}",
    eprint = "2503.00636",
    archivePrefix = "arXiv",
    primaryClass = "astro-ph.IM",
    reportNumber = "FERMILAB-PUB-25-0188-PPD",
    doi = "10.1088/1475-7516/2025/08/034",
    journal = "JCAP",
    volume = "08",
    pages = "034",
    year = "2025"
}

@article{Kacprzak:2022pww,
    author = "Kacprzak, Tomasz and Fluri, Janis and Schneider, Aurel and Refregier, Alexandre and Stadel, Joachim",
    title = "{CosmoGridV1: a simulated {\ensuremath{\mathsf{w}}}CDM theory prediction for map-level cosmological inference}",
    eprint = "2209.04662",
    archivePrefix = "arXiv",
    primaryClass = "astro-ph.CO",
    doi = "10.1088/1475-7516/2023/02/050",
    journal = "JCAP",
    volume = "02",
    pages = "050",
    year = "2023"
}

@article{Fluri:2022rvb,
    author = "Fluri, Janis and Kacprzak, Tomasz and Lucchi, Aurelien and Schneider, Aurel and Refregier, Alexandre and Hofmann, Thomas",
    title = "{Full wCDM analysis of KiDS-1000 weak lensing maps using deep learning}",
    eprint = "2201.07771",
    archivePrefix = "arXiv",
    primaryClass = "astro-ph.CO",
    doi = "10.1103/PhysRevD.105.083518",
    journal = "Phys. Rev. D",
    volume = "105",
    number = "8",
    pages = "083518",
    year = "2022"
}

@article{Dietrich:2009jq,
    author = "Dietrich, J. P. and Hartlap, J.",
    title = "{Cosmology with the shear-peak statistics}",
    eprint = "0906.3512",
    archivePrefix = "arXiv",
    primaryClass = "astro-ph.CO",
    doi = "10.1111/j.1365-2966.2009.15948.x",
    journal = "Mon. Not. Roy. Astron. Soc.",
    volume = "402",
    pages = "1049",
    year = "2010"
}

@article{Heitmann:2008eq,
    author = "Heitmann, Katrin and White, Martin and Wagner, Christian and Habib, Salman and Higdon, David",
    title = "{The Coyote Universe I: Precision Determination of the Nonlinear Matter Power Spectrum}",
    eprint = "0812.1052",
    archivePrefix = "arXiv",
    primaryClass = "astro-ph",
    reportNumber = "LA-UR-08-05630",
    doi = "10.1088/0004-637X/715/1/104",
    journal = "Astrophys. J.",
    volume = "715",
    pages = "104--121",
    year = "2010"
}

@article{Heitmann:2009cu,
    author = "Heitmann, Katrin and Higdon, David and White, Martin and Habib, Salman and Williams, Brian J. and Wagner, Christian",
    title = "{The Coyote Universe II: Cosmological Models and Precision Emulation of the Nonlinear Matter Power Spectrum}",
    eprint = "0902.0429",
    archivePrefix = "arXiv",
    primaryClass = "astro-ph.CO",
    reportNumber = "LA-UR-08-07921",
    doi = "10.1088/0004-637X/705/1/156",
    journal = "Astrophys. J.",
    volume = "705",
    pages = "156--174",
    year = "2009"
}

@article{Lawrence:2009uk,
    author = "Lawrence, Earl and Heitmann, Katrin and White, Martin and Higdon, David and Wagner, Christian and Habib, Salman and Williams, Brian",
    title = "{The Coyote Universe III: Simulation Suite and Precision Emulator for the Nonlinear Matter Power Spectrum}",
    eprint = "0912.4490",
    archivePrefix = "arXiv",
    primaryClass = "astro-ph.CO",
    reportNumber = "LA-UR-09-06131",
    doi = "10.1088/0004-637X/713/2/1322",
    journal = "Astrophys. J.",
    volume = "713",
    pages = "1322--1331",
    year = "2010"
}

@article{ZorrillaMatilla:2016vae,
    author = "Zorrilla Matilla, Jos{\'e} Manuel and Haiman, Zolt{\'a}n and Hsu, Daniel and Gupta, Arushi and Petri, Andrea",
    title = "{Do dark matter halos explain lensing peaks?}",
    eprint = "1609.03973",
    archivePrefix = "arXiv",
    primaryClass = "astro-ph.CO",
    doi = "10.1103/PhysRevD.94.083506",
    journal = "Phys. Rev. D",
    volume = "94",
    number = "8",
    pages = "083506",
    year = "2016"
}

@article{Gupta:2018eev,
    author = "Gupta, Arushi and Matilla, Jos{\'e} Manuel Zorrilla and Hsu, Daniel and Haiman, Zolt{\'a}n",
    title = "{Non-Gaussian information from weak lensing data via deep learning}",
    eprint = "1802.01212",
    archivePrefix = "arXiv",
    primaryClass = "astro-ph.CO",
    doi = "10.1103/PhysRevD.97.103515",
    journal = "Phys. Rev. D",
    volume = "97",
    number = "10",
    pages = "103515",
    year = "2018"
}

@article{Lesgourgues:2011re,
    author = "Lesgourgues, Julien",
    title = "{The Cosmic Linear Anisotropy Solving System (CLASS) I: Overview}",
    eprint = "1104.2932",
    archivePrefix = "arXiv",
    primaryClass = "astro-ph.IM",
    month = "4",
    year = "2011"
}

@article{Lewis:1999bs,
    author = "Lewis, Antony and Challinor, Anthony and Lasenby, Anthony",
    title = "{Efficient computation of CMB anisotropies in closed FRW models}",
    eprint = "astro-ph/9911177",
    archivePrefix = "arXiv",
    doi = "10.1086/309179",
    journal = "Astrophys. J.",
    volume = "538",
    pages = "473--476",
    year = "2000"
}

@article{Wright:2025xka,
    author = "Wright, Angus H. and others",
    title = "{KiDS-Legacy: Cosmological constraints from cosmic shear with the complete Kilo-Degree Survey}",
    eprint = "2503.19441",
    archivePrefix = "arXiv",
    primaryClass = "astro-ph.CO",
    doi = "10.1051/0004-6361/202554908",
    journal = "Astron. Astrophys.",
    volume = "703",
    pages = "A158",
    year = "2025"
}

@article{DES:2026mkc,
    author = "Abbott, T. M. C. and others",
    collaboration = "DES",
    title = "{Dark Energy Survey Year 6 Results: Cosmological Constraints from Cosmic Shear}",
    eprint = "2602.10065",
    archivePrefix = "arXiv",
    primaryClass = "astro-ph.CO",
    reportNumber = "FERMILAB-PUB-26-0027-PPD",
    month = "2",
    year = "2026"
}

@article{Miyatake:2023njf,
    author = "Miyatake, Hironao and others",
    title = "{Hyper Suprime-Cam Year 3 results: Cosmology from galaxy clustering and weak lensing with HSC and SDSS using the emulator based halo model}",
    eprint = "2304.00704",
    archivePrefix = "arXiv",
    primaryClass = "astro-ph.CO",
    doi = "10.1103/PhysRevD.108.123517",
    journal = "Phys. Rev. D",
    volume = "108",
    number = "12",
    pages = "123517",
    year = "2023"
}

@article{Amendola:2016saw,
    author = "Amendola, Luca and others",
    title = "{Cosmology and fundamental physics with the Euclid satellite}",
    eprint = "1606.00180",
    archivePrefix = "arXiv",
    primaryClass = "astro-ph.CO",
    doi = "10.1007/s41114-017-0010-3",
    journal = "Living Rev. Rel.",
    volume = "21",
    number = "1",
    pages = "2",
    year = "2018"
}

@article{Akeson:2019biv,
    author = "Akeson, Rachel and others",
    title = "{The Wide Field Infrared Survey Telescope: 100 Hubbles for the 2020s}",
    eprint = "1902.05569",
    archivePrefix = "arXiv",
    primaryClass = "astro-ph.IM",
    month = "2",
    year = "2019"
}

@article{Spergel:2015sza,
    author = "Spergel, D. and others",
    title = "{Wide-Field InfrarRed Survey Telescope-Astrophysics Focused Telescope Assets WFIRST-AFTA 2015 Report}",
    eprint = "1503.03757",
    archivePrefix = "arXiv",
    primaryClass = "astro-ph.IM",
    month = "3",
    year = "2015"
}

@article{Takahashi:2012em,
    author = "Takahashi, Ryuichi and Sato, Masanori and Nishimichi, Takahiro and Taruya, Atsushi and Oguri, Masamune",
    title = "{Revising the Halofit Model for the Nonlinear Matter Power Spectrum}",
    eprint = "1208.2701",
    archivePrefix = "arXiv",
    primaryClass = "astro-ph.CO",
    doi = "10.1088/0004-637X/761/2/152",
    journal = "Astrophys. J.",
    volume = "761",
    pages = "152",
    year = "2012"
}

@article{Minkowski1903,
    author = {Minkowski, Hermann},
    journal = {Mathematische Annalen},
    pages = {447-495},
    title = {Volumen und Oberfläche},
    url = {http://eudml.org/doc/158108},
    volume = {57},
    year = {1903},
}

@article{Bardeen:1985tr,
    author = "Bardeen, James M. and Bond, J. R. and Kaiser, Nick and Szalay, A. S.",
    title = "{The Statistics of Peaks of Gaussian Random Fields}",
    reportNumber = "FERMILAB-PUB-85-148-A, NSF-ITP-85-93",
    doi = "10.1086/164143",
    journal = "Astrophys. J.",
    volume = "304",
    pages = "15--61",
    year = "1986"
}

@ARTICLE{1987MNRAS.226..655B,
       author = {{Bond}, J.~R. and {Efstathiou}, G.},
        title = "{The statistics of cosmic background radiation fluctuations}",
      journal = {MNRAS},
         year = 1987,
        month = jun,
       volume = {226},
        pages = {655-687},
          doi = {10.1093/mnras/226.3.655},
       adsurl = {https://ui.adsabs.harvard.edu/abs/1987MNRAS.226..655B}
}

@ARTICLE{2011arXiv1101.2286M,
       author = {{Mallat}, St{\'e}phane},
        title = "{Group Invariant Scattering}",
      journal = {arXiv e-prints},
         year = 2011,
        month = jan,
          eid = {arXiv:1101.2286},
        pages = {arXiv:1101.2286},
          doi = {10.48550/arXiv.1101.2286},
archivePrefix = {arXiv},
       eprint = {1101.2286},
 primaryClass = {math.FA},
       adsurl = {https://ui.adsabs.harvard.edu/abs/2011arXiv1101.2286M}
}

@article{Bird:2011rb,
    author = "Bird, Simeon and Viel, Matteo and Haehnelt, Martin G.",
    title = "{Massive Neutrinos and the Non-linear Matter Power Spectrum}",
    eprint = "1109.4416",
    archivePrefix = "arXiv",
    primaryClass = "astro-ph.CO",
    doi = "10.1111/j.1365-2966.2011.20222.x",
    journal = "Mon. Not. Roy. Astron. Soc.",
    volume = "420",
    pages = "2551--2561",
    year = "2012"
}

@article{Bayer:2024egi,
    author = "Bayer, Adrian E. and Zhong, Yici and Li, Zack and DeRose, Joseph and Feng, Yu and Liu, Jia",
    title = "{The HalfDome multi-survey cosmological simulations: N-body simulations}",
    eprint = "2407.17462",
    archivePrefix = "arXiv",
    primaryClass = "astro-ph.CO",
    doi = "10.1088/1475-7516/2025/05/016",
    journal = "JCAP",
    volume = "05",
    pages = "016",
    year = "2025"
}

@article{Takahashi:2017hjr,
    author = "Takahashi, Ryuichi and Hamana, Takashi and Shirasaki, Masato and Namikawa, Toshiya and Nishimichi, Takahiro and Osato, Ken and Shiroyama, Kosei",
    title = "{Full-sky Gravitational Lensing Simulation for Large-area Galaxy Surveys and Cosmic Microwave Background Experiments}",
    eprint = "1706.01472",
    archivePrefix = "arXiv",
    primaryClass = "astro-ph.CO",
    doi = "10.3847/1538-4357/aa943d",
    journal = "Astrophys. J.",
    volume = "850",
    number = "1",
    pages = "24",
    year = "2017"
}

@article{Anbajagane:2023wif,
    author = "Anbajagane, Dhayaa and Chang, Chihway and Lee, Hayden and Gatti, Marco",
    title = "{Primordial non-Gaussianities with weak lensing: information on non-linear scales in the Ulagam full-sky simulations}",
    eprint = "2310.02349",
    archivePrefix = "arXiv",
    primaryClass = "astro-ph.CO",
    doi = "10.1088/1475-7516/2024/03/062",
    journal = "JCAP",
    volume = "03",
    pages = "062",
    year = "2024"
}

@article{DES:2026fyc,
    author = "Abbott, T. M. C. and others",
    collaboration = "DES",
    title = "{Dark Energy Survey Year 6 Results: Cosmological Constraints from Galaxy Clustering and Weak Lensing}",
    eprint = "2601.14559",
    archivePrefix = "arXiv",
    primaryClass = "astro-ph.CO",
    reportNumber = "DES-2025-0929, FERMILAB-PUB-26-0026-PPD",
    month = "1",
    year = "2026"
}

@article{ACT:2023dou,
    author = "Qu, Frank J. and others",
    collaboration = "ACT",
    title = "{The Atacama Cosmology Telescope: A Measurement of the DR6 CMB Lensing Power Spectrum and Its Implications for Structure Growth}",
    eprint = "2304.05202",
    archivePrefix = "arXiv",
    primaryClass = "astro-ph.CO",
    reportNumber = "FERMILAB-PUB-23-237-PPD, FERMILAB-PUB-23-237-PPD",
    doi = "10.3847/1538-4357/acfe06",
    journal = "Astrophys. J.",
    volume = "962",
    number = "2",
    pages = "112",
    year = "2024"
}

@article{ACT:2023kun,
    author = "Madhavacheril, Mathew S. and others",
    collaboration = "ACT",
    title = "{The Atacama Cosmology Telescope: DR6 Gravitational Lensing Map and Cosmological Parameters}",
    eprint = "2304.05203",
    archivePrefix = "arXiv",
    primaryClass = "astro-ph.CO",
    reportNumber = "FERMILAB-PUB-23-206-PPD",
    doi = "10.3847/1538-4357/acff5f",
    journal = "Astrophys. J.",
    volume = "962",
    number = "2",
    pages = "113",
    year = "2024"
}

@article{Wu:2019hek,
    author = "Wu, W. L. K. and others",
    title = "{A Measurement of the Cosmic Microwave Background Lensing Potential and Power Spectrum from 500 deg$^2$ of SPTpol Temperature and Polarization Data}",
    eprint = "1905.05777",
    archivePrefix = "arXiv",
    primaryClass = "astro-ph.CO",
    doi = "10.3847/1538-4357/ab4186",
    journal = "Astrophys. J.",
    volume = "884",
    pages = "70",
    year = "2019"
}

@article{SPT:2023jql,
    author = "Pan, Z. and others",
    collaboration = "SPT",
    title = "{Measurement of gravitational lensing of the cosmic microwave background using SPT-3G 2018 data}",
    eprint = "2308.11608",
    archivePrefix = "arXiv",
    primaryClass = "astro-ph.CO",
    doi = "10.1103/PhysRevD.108.122005",
    journal = "Phys. Rev. D",
    volume = "108",
    number = "12",
    pages = "122005",
    year = "2023"
}

@article{Planck:2018lbu,
    author = "Aghanim, N. and others",
    collaboration = "Planck",
    title = "{Planck 2018 results. VIII. Gravitational lensing}",
    eprint = "1807.06210",
    archivePrefix = "arXiv",
    primaryClass = "astro-ph.CO",
    doi = "10.1051/0004-6361/201833886",
    journal = "Astron. Astrophys.",
    volume = "641",
    pages = "A8",
    year = "2020"
}

@article{Carron:2022eyg,
    author = "Carron, Julien and Mirmelstein, Mark and Lewis, Antony",
    title = "{CMB lensing from Planck PR4~maps}",
    eprint = "2206.07773",
    archivePrefix = "arXiv",
    primaryClass = "astro-ph.CO",
    doi = "10.1088/1475-7516/2022/09/039",
    journal = "JCAP",
    volume = "09",
    pages = "039",
    year = "2022"
}

@article{Chen:2026jlr,
    author = "Chen, Shu-Fan and Hill, J. Colin and Haiman, Zolt{\'a}n",
    title = "{Probing Cosmology through Higher-Order CMB Lensing Statistics}",
    eprint = "2603.12335",
    archivePrefix = "arXiv",
    primaryClass = "astro-ph.CO",
    month = "3",
    year = "2026"
}

@article{Liu:2016nfs,
    author = {Liu, Jia and Hill, J. Colin and Sherwin, Blake D. and Petri, Andrea and B{\"o}hm, Vanessa and Haiman, Zolt{\'a}n},
    title = "{CMB lensing beyond the power spectrum: Cosmological constraints from the one-point probability distribution function and peak counts}",
    eprint = "1608.03169",
    archivePrefix = "arXiv",
    primaryClass = "astro-ph.CO",
    doi = "10.1103/PhysRevD.94.103501",
    journal = "Phys. Rev. D",
    volume = "94",
    number = "10",
    pages = "103501",
    year = "2016"
}

@article{Kratochvil:2009wh,
    author = "Kratochvil, Jan M. and Haiman, Zoltan and May, Morgan",
    title = "{Probing Cosmology with Weak Lensing Peak Counts}",
    eprint = "0907.0486",
    archivePrefix = "arXiv",
    primaryClass = "astro-ph.CO",
    doi = "10.1103/PhysRevD.81.043519",
    journal = "Phys. Rev. D",
    volume = "81",
    pages = "043519",
    year = "2010"
}
\end{document}